\documentclass[twocolumn,useAMS,usenatbib]{mn2e}

\usepackage{graphicx}
\usepackage{subfigure}
\usepackage{xcolor}
\usepackage{mathtext,bbm,amsmath,amsfonts,amssymb,indentfirst,syntonly,graphicx}
\usepackage{bm}
\usepackage{mathtools}
\usepackage{slashbox}
\usepackage[english]{babel}
\usepackage{calc}
\usepackage{tikz}

\def\bc{\begin{center}}
\def\ec{\end{center}}
\def\be{\begin{eqnarray}}
\def\ee{\end{eqnarray}}

\title[The significance of anisotropic signals]{The significance of anisotropic signals hiding in the type Ia supernovae}
\author[H.-N. Lin, X. Li and Z. Chang]
        {Hai-Nan Lin$^{1}$\thanks{linhn@ihep.ac.cn}, Xin Li$^{1,2}$\thanks{lixin1981@cqu.edu.cn} and Zhe Chang$^{3}$\\
$^{1}$Department of Physics, Chongqing University, Chongqing 401331, China\\
$^{2}$State Key Laboratory of Theoretical Physics, Institute of Theoretical Physics, Chinese Academy of Sciences, Beijing 100190, China\\
$^{3}$Institute of High Energy Physics, Chinese Academy of Sciences, Beijing 100049, China}

\begin{document}

\date{Accepted xxxx; Received xxxx; in original form xxxx}

\pagerange{\pageref{firstpage}--\pageref{lastpage}} \pubyear{2016}

\maketitle

\label{firstpage}

\begin{abstract}
  We use two different methods, i.e., dipole-fitting (DF) and hemisphere comparison (HC), to search for the anisotropic signals hiding in the Union2.1 data set. We find that the directions of maximum matter density derived using these two methods are about $114^{\circ}$ away from each other. We construct four Union2.1-like mock samples to test the statistical significance of these two methods. It is shown that DF method is statistically significant, while HC method is strongly biased by the distribution of data points in the sky. Then we assume that the anisotropy of distance modulus is mainly induced by the anisotropy of matter density, which is modeled to be the dipole form $\Omega_M=\Omega_{M0}(1-\cos\theta)$. We fit our model to Union2.1, and find that the direction of maximum matter density is well consistent with the direction derived using DF method, but it is very different from the direction previously claimed. Monte Carlo simulations show that our method is statistically more significant than HC method, although it is not as significant as DF method. The statistical significance can be further improved if the data points are homogeneously distributed in the sky. Due to the low quality of present supernovae data, however, it is still premature to claim that the Universe has any preferred direction.
\end{abstract}


\section{Introduction}\label{sec:introduction}

Progresses in both the theoretical and observational aspects in past decades have lead to the developments of the standard cosmological model, i.e., the cold dark matter plus a cosmological constant ($\Lambda$CDM) model. One of the foundations of $\Lambda$CDM model is the so-called cosmological principle, which says that the Universe is homogeneous and isotropic on large scale. The $\Lambda$CDM model is to some degree in accordance with cosmological observations, such as the statistics of galaxies \citep{Trujillo-Gomez:2011}, the halo power spectrum \citep{Reid:2010}. Particularly, the approximate isotropy of the cosmic microwave background (CMB) radiation from the {\it Wilkinson Microwave Anisotropy Probe} ({\it WMAP}) \citep{Bennett:2013,Hinshaw:2013} and {\it Planck} \citep{Planck:2014a,Planck:2015a} satellites is a strong support to the cosmological principle. The $\Lambda$CDM model has succeeded in accounting for some properties of the Universe, e.g., the large-scale structure in the distribution of galaxies, the abundances of hydrogen and helium, the accelerating expansion of the Universe, and so on.

Although $\Lambda$CDM model has achieved great successes, it also confronts many challenges. The main difficulty that $\Lambda$CDM model encounters is the explanation for some anisotropic phenomena. These include the alignments of low multipoles in the CMB angular power spectrum \citep{Lineweaver:1996,Tegmark:2003,Bielewicz:2004,Copi:2010,Frommert:2010}, the spatial variation of the fine-structure constant \citep{Webb:2011,King:2012,Mariano:2012}, the large-scale alignments of quasar polarization vectors \citep{Hutsemekers:2005,Hutsemekers:2011}, the unexpected large-scale bulk flow \citep{Kashlinsky:2008,Watkins:2009,Lavaux:2010}. Especially, the hemispherical asymmetry of CMB \citep{Eriksen:2004,Akrami:2014,Planck:2014b,Planck:2015b} puts a serious challenge on $\Lambda$CDM model. These imply that the Universe may deviate from isotropy and the validity of cosmological principle is questionable. In the theoretical aspect, some anisotropic cosmological models have been proposed, such as the Bianchi type I cosmological model \citep{Campanelli:2011,Schcker:2014}, the extended topological quintessence model \citep{Mariano:2012}, the $\Lambda$CDM model with a scalar perturbation \citep{Li:2013}, the Finsler-Randers cosmological model \citep{Chang:2013,Chang:2014a,Chang:2014b,Li:2015}.

At the end of the twentieth century, two collaborations independently found by measuring the luminosity distance of type Ia supernovae (SNe Ia) that the Universe is accelerated expanding \citep{Riess:1998,Perlmutter:1999}. Since then, SNe Ia are widely used as the distance indicators to trace the expansion of the Universe. This is because SNe Ia are often assumed to be produced by the gravitational collapse of white dwarfs whose mass exceeds the Chandrasekhar limit, hence all SNe Ia have approximately the same absolute magnitude. Until now, SNe Ia are still among the most ideal standard candles to probe the large-scale structure of the Universe. Recently, some methods have been developed to single out the anisotropic signals hiding in the SNe Ia data. Assuming the observed distance modulus deviating from the theoretical prediction of $\Lambda$CDM model is the dipole form, it was shown that the dipole component is necessary at more than $2\sigma$ confidence level \citep{Mariano:2012,Cai:2013,Yang:2014,Wang:2014}. By dividing the sky into two opposite hemispheres and comparing the cosmological parameters (such as $\Omega_M$, $H_0$ and $q_0$) in each hemisphere, some preferred directions have been identified \citep{Antoniou:2010,CaiTuo:2012,kalus:2013,Yang:2014}. A similar method, except that the hemisphere is replaced by a spherical cap of fixed stretch angle, has been developed \citep{Bengaly:2015,Javanmardi:2015}. Another way is to divide the sky into $N$ patches of the same area, and compare the cosmological parameters in each patch \citep{Zhao:2013,Carvalho:2015}. A method called smoothed residual has been used to search for the preferred direction \citep{Colin:2011,Feindt:2013,Appleby:2014}.

Unfortunately, all the SNe Ia data sets only cover a limited region of the full sky. In other words, the angular distribution of data points in the sky is far from homogeneous. On the other hand, the systematic uncertainties are still large. It is doubtable whether the anisotropic signals are arising from the intrinsic property of the Universe, or just due to the statistical noise. \citet{Bengaly:2015} showed that the celestial incompleteness of current SNe Ia samples may lead to the anisotropic signals. It was noted that the preferred directions derived from the Union2 \citep{Amanullah:2010} data set using two different methods, i.e., dipole-fitting (DF) method and hemisphere comparison (HC) method, are approximately opposite \citep{Chang:2015}. It was shown that the discrepancy may be due to the inhomogeneous angular distribution of the data points. Moreover, \citet{Lin:2016} found the dipole direction of the JLA \citep{Betoule:2014} data set is approximately opposite to that of the Union2 data set. In general, the anisotropic signals singled out from SNe Ia depend on both the methods and data sets that are used in the analysis. Therefore, it is still premature to make a conclusive conclusion at present.

This paper is the continuation of our previous work \citep{Chang:2015}. The aim of this paper is to test the statistical significance of anisotropic signals hiding in the supernovae data sets. We first use two different methods, i.e., DF and HC, to search for the preferred direction in the Union2.1 \citep{Suzuki:2012}. We will show that the preferred directions obtained using these two methods are very different. Then we use Monte Carlo (MC) simulations to show that DF method is statistically significant. However, HC method couldn't correctly reproduce the fiducial direction when the data points are really anisotropic. Moreover, HC method may pick out pseudo anisotropic signals when the data points are actually isotropic. To alleviate the discrepancy between these two methods, we model the matter density to be the dipole form. We fit our model to Union2.1, and obtained the preferred direction which is consistent with the result of DF method.

The rest of the paper is arranged as follows: In section \ref{sec:dipole}, we briefly introduce the DF method, then use it to search for the dipole direction of Union2.1. MC simulations are applied to test the statistical significance. In section \ref{sec:hemisphere}, we exploit the HC method to search for the preferred direction hiding in Union2.1, and test the statistical significance using MC simulations. In section \ref{sec:matterdensity}, we parameterize the matter density of the Universe to be the dipole form, and search for the direction of maximum matter density. Finally, discussions and and conclusions are given in section \ref{sec:discussion}.

\section{Dipole-fitting}\label{sec:dipole}

In this section, we first briefly introduce the DF method and the Union2.1 data set that are used in our analysis. Then we use DF method to search for the possible anisotropy of Union2.1. Finally, we apply MC simulations to test the statistical significance of our results. In what follows, we assume that the isotropic background space time is described by the spatially flat $\Lambda$CDM model. In this model, the distance-redshift relation is given by
\begin{equation}\label{eq:lumi_distance}
  \bar{d}_L(z)=(1+z)\frac{c}{H_0}\int_0^z\frac{dz}{\sqrt{\Omega_M(1+z)^3+(1-\Omega_M)}},
\end{equation}
where $H_0$ is the Hubble constant today, and $\Omega_M$ is the normalized matter density. The distance modulus is related to the luminosity distance by
\begin{equation}\label{eq:distance_modulus}
  \bar{\mu}(z)=5\log\frac{\bar{d}_L(z)}{\rm Mpc}+25.
\end{equation}
Throughout this paper, a quantity with a bar over it means that it is calculated from $\Lambda$CDM model.

The parameters $H_0$ and $\Omega_M$ are derived from fitting to the Union2.1 data set. The Union2.1 is a compilation of 580 well calibrated SNe Ia in the redshift range $z\in[0.015,1.414]$ \citep{Suzuki:2012}. Each supernova has well measured redshift $z$, distance modulus $\mu$ and its uncertainty $\sigma_{\mu}$, and the position in the equatorial coordinates. To directly compare with previous works, we transform the equatorial coordinates into galactic coordinates. The best-fitting parameters are the ones which can minimize $\chi^2$,
\begin{equation}\label{eq:chi2}
  \chi^2=\sum_{i=1}^{580}\left[\frac{\mu_{\rm th}(z_i)-\mu_{{\rm obs},i}}{\sigma_{\mu,i}}\right]^2,
\end{equation}
where $\mu_{\rm th}$ is the theoretical distance modulus calculated using equations (\ref{eq:lumi_distance}) and (\ref{eq:distance_modulus}). The result is
\begin{equation}\label{eq:parameter_iso}
  \Omega_M=0.278\pm 0.019,~~ H_0=70.0\pm 0.3~{\rm km~s}^{-1}~{\rm Mpc}^{-1}.
\end{equation}
Here and follows, all the uncertainties are given at $1\sigma$ ($68\%$) confidence level.

Next, we assume the ``\,real" distance modulus is not isotropic, but is the dipole form, i.e.,
\begin{equation}\label{eq:mu_dipole}
  \mu(z)=\bar{\mu}(z)[1-D(\hat{\bm n}\cdot\hat{\bm p})],
\end{equation}
where $D$ is the dipole amplitude, $\hat{\bm n}$ and $\hat{\bm p}$ are the unit vectors pointing towards the dipole direction and supernova position, respectively. In the galactic coordinates, the dipole direction can be parameterized as $\hat{\bm n}=\cos(b)\cos(l)\hat{\bm i}+\cos(b)\sin(l)\hat{\bm j}+\sin(b)\hat{\bm k}$, where $l\in[0^{\circ},360^{\circ})$ and $b\in[-90^{\circ},+90^{\circ}]$ are the galactic longitude and latitude, respectively, and $\hat{\bm i}$, $\hat{\bm j}$, $\hat{\bm k}$ are three orthogonal unit vectors. Similarly, the position of the $i$th supernova can be written as $\hat{\bm p}_i=\cos(b_i)\cos(l_i)\hat{\bm i}+\cos(b_i)\sin(l_i)\hat{\bm j}+\sin(b_i)\hat{\bm k}$.

Now the theoretical distance modulus in equation (\ref{eq:chi2}) is given by equation (\ref{eq:mu_dipole}). Minimizing the right-hand-side of equation (\ref{eq:chi2}) leads to the dipole amplitude
\begin{equation}\label{eq:dipole_amplitude}
  D=(1.2\pm 0.5) \times 10^{-3},
\end{equation}
and the dipole direction
\begin{equation}\label{dipole_direction}
  (l,b)=(310.6^{\circ}\pm 18.2^{\circ}, -13.0^{\circ}\pm 11.1^{\circ}).
\end{equation}
This result is obtained from fixing $\Omega_M$ and $H_0$ to the values given in equation (\ref{eq:parameter_iso}). Equation (\ref{eq:dipole_amplitude}) implies that the Union2.1 data set deviates from isotropy at more than $2\sigma$ confidence level. The dipole direction in equation (\ref{dipole_direction}) is consistent with that obtained by \citet{Yang:2014} using the same data set. Both the dipole amplitude and dipole direction are well consistent with that derived from Union2 \citep{Mariano:2012}.

The distribution of supernovae in the sky is extremely inhomogeneous. Due to the narrow sky coverage of detectors, especially the Sloan Digital Sky Survey (SDSS) supernovae survey, more than one half data points cluster near the celestial equator, with declination smaller than $2.5^{\circ}$. Besides, the measured distance moduli extracted from light curves have an average uncertainty at the order of $0.2$ mag. To check if the inhomogeneous distribution of data points have some influences on our results, and also to check if the anisotropic signals are indeed arising from the intrinsic property of the Universe or just due to the statistical noise, we create four Union2.1-like mock samples to test the statistical significance our results.
\begin{itemize}
  \item{Sample A: The positions are unchanged, and the distance moduli have a fiducial dipole direction. This can be done by replacing the distance modulus of the $i$th supernova with a random number generated from the Gaussian distribution $G(\mu_i, \sigma_{\mu_i})$, where $\mu_i=\bar{\mu}(z_i)[1-D(\bm{\hat{n}}\cdot\bm{\hat{p}}_i)]$, $D$ and $\bm{\hat{n}}$ are respectively the dipole amplitude and dipole direction of Union2.1, which are given in equations (\ref{eq:dipole_amplitude}) and (\ref{dipole_direction}). The redshift $z_i$, position $(l_i,b_i)$, and the uncertainty of distance modulus $\sigma_{\mu_i}$ are reproduced from Union2.1.}
  \item{Sample B: The positions are homogeneously distributed in the sky, and the distance moduli have a fiducial dipole direction. This can be done by replacing the coordinates of the $i$th supernova $(l_i, b_i)$ with a pair of random numbers following the uniform distribution, and replacing the distance modulus with a random number generated from $G(\mu_i, \sigma_{\mu_i})$. The redshift $z_i$ and the uncertainty of distance modulus $\sigma_{\mu_i}$ are reproduced from Union2.1.}
  \item{Sample C: The positions are unchanged, and the distance moduli are fiducially isotropic. This can be done by replacing the distance modulus of the $i$th supernova with a random number generated from the Gaussian distribution $G(\bar{\mu}_i, \sigma_{\mu_i})$, where $\bar{\mu}_i$ is the isotropic distance modulus predicted by $\Lambda$CDM model. All the rest observables are reproduced from Union2.1.}
  \item{Sample D: The positions are homogeneously distributed in the sky, and the distance moduli are fiducially isotropic. This can be done by replacing the coordinates of the $i$th supernova $(l_i, b_i)$ with a pair of random numbers following the uniform distribution, and replacing the distance modulus with a random number generated from $G(\bar{\mu}_i, \sigma_{\mu_i})$. The redshift $z_i$ and the uncertainty of distance modulus $\sigma_{\mu_i}$ are reproduced from Union2.1.}
\end{itemize}

\begin{figure}
  \centering
  \includegraphics[width=0.5\textwidth]{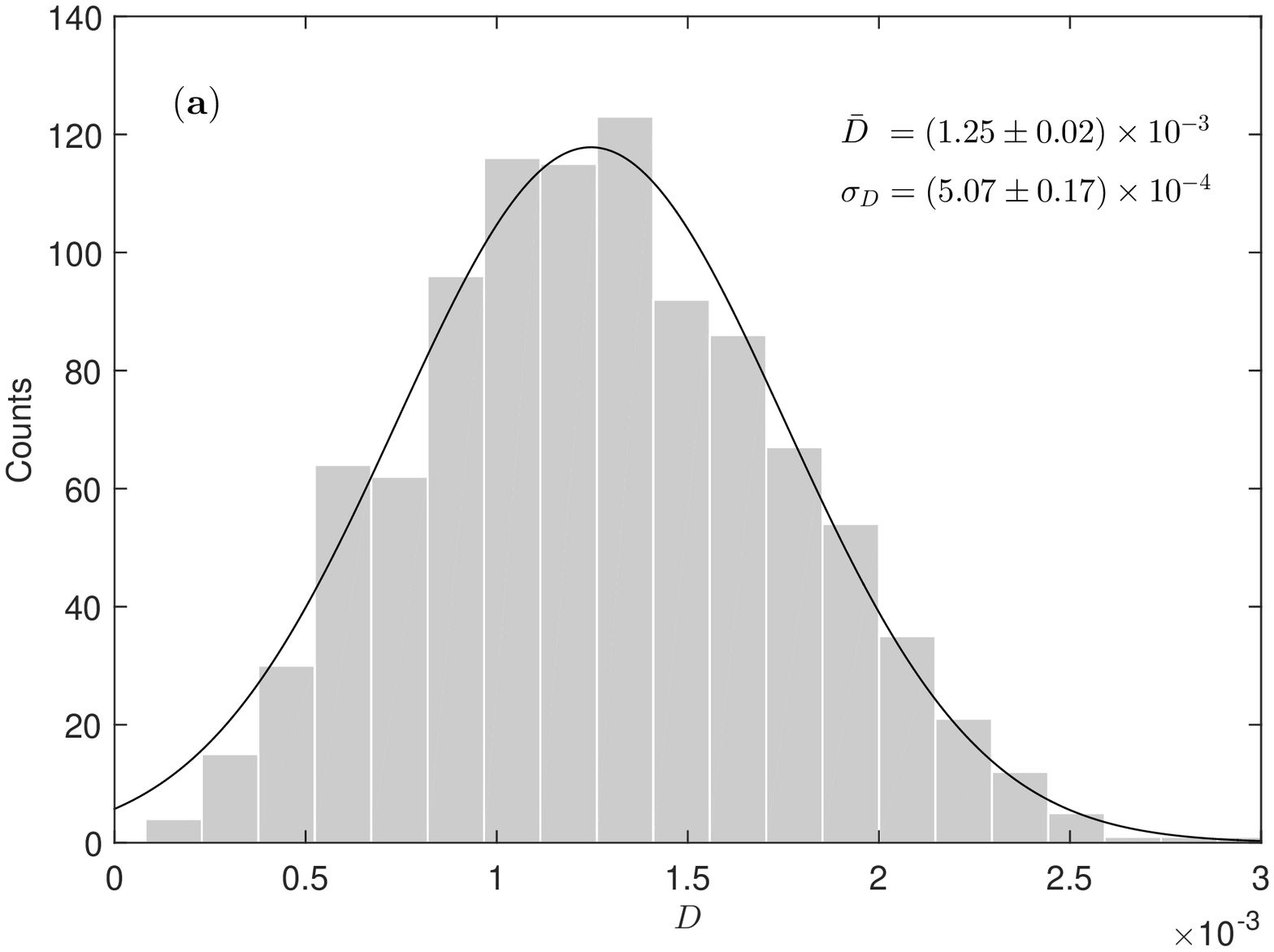}
  \includegraphics[width=0.5\textwidth]{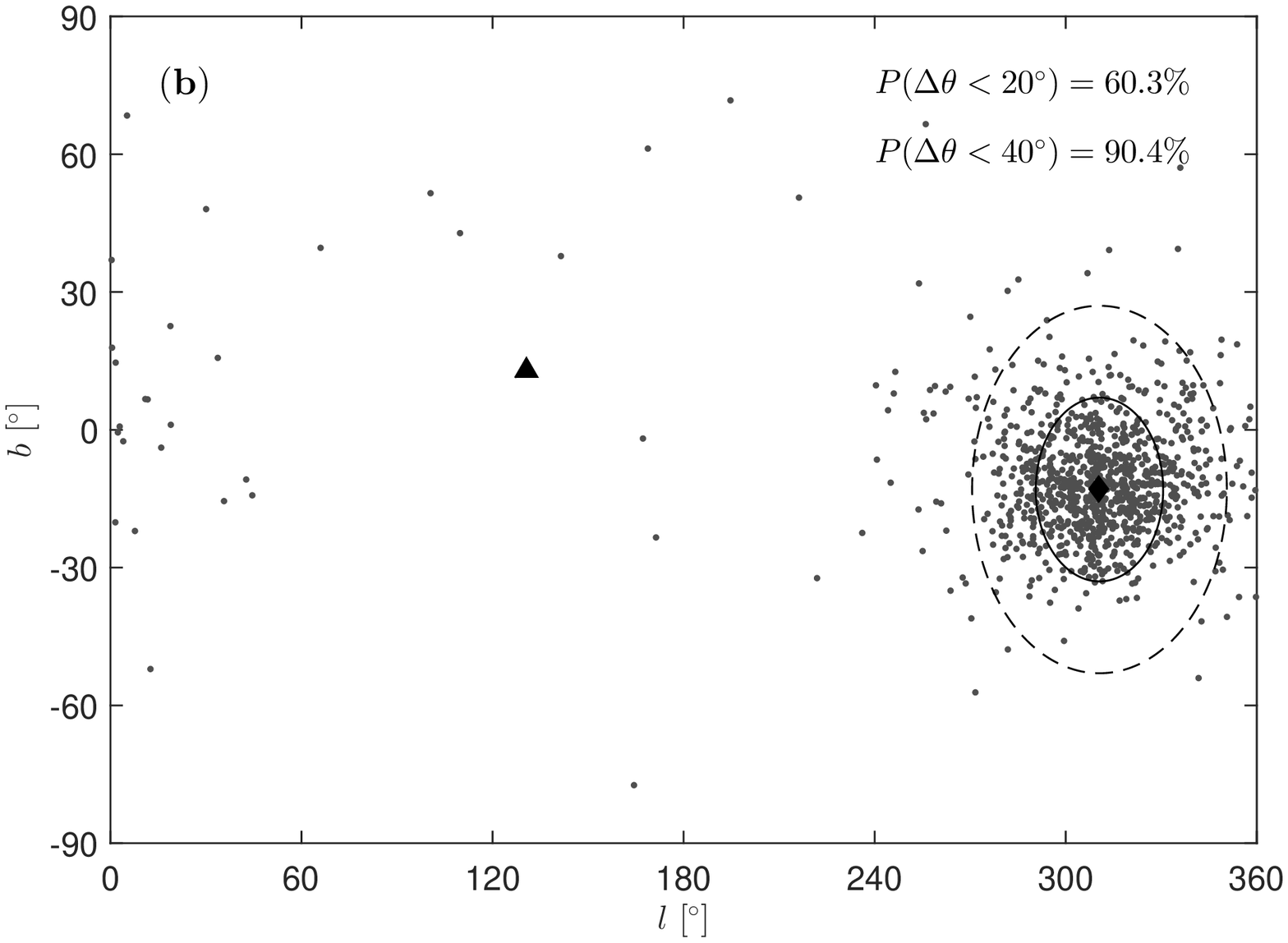}
  \caption{The dipoles of mock sample A in 1000 simulations. Panel (a): the histogram of dipole amplitudes, with black curve the best-fitting result to Gaussian function. Panel (b): the dipole directions in the sky of galactic coordinates. The black diamond is the fiducial dipole direction pointing towards $(l,b)=(310.6^{\circ},-13.0^{\circ})$, and the black triangle is the antipode. The solid (dashed) circle represents a circular region of radius $\Delta\theta<20^{\circ}$ ($\Delta\theta<40^{\circ}$), centering on the fiducial dipole direction.}
  \label{fig:DF1}
\end{figure}

\begin{figure}
  \centering
  \includegraphics[width=0.5\textwidth]{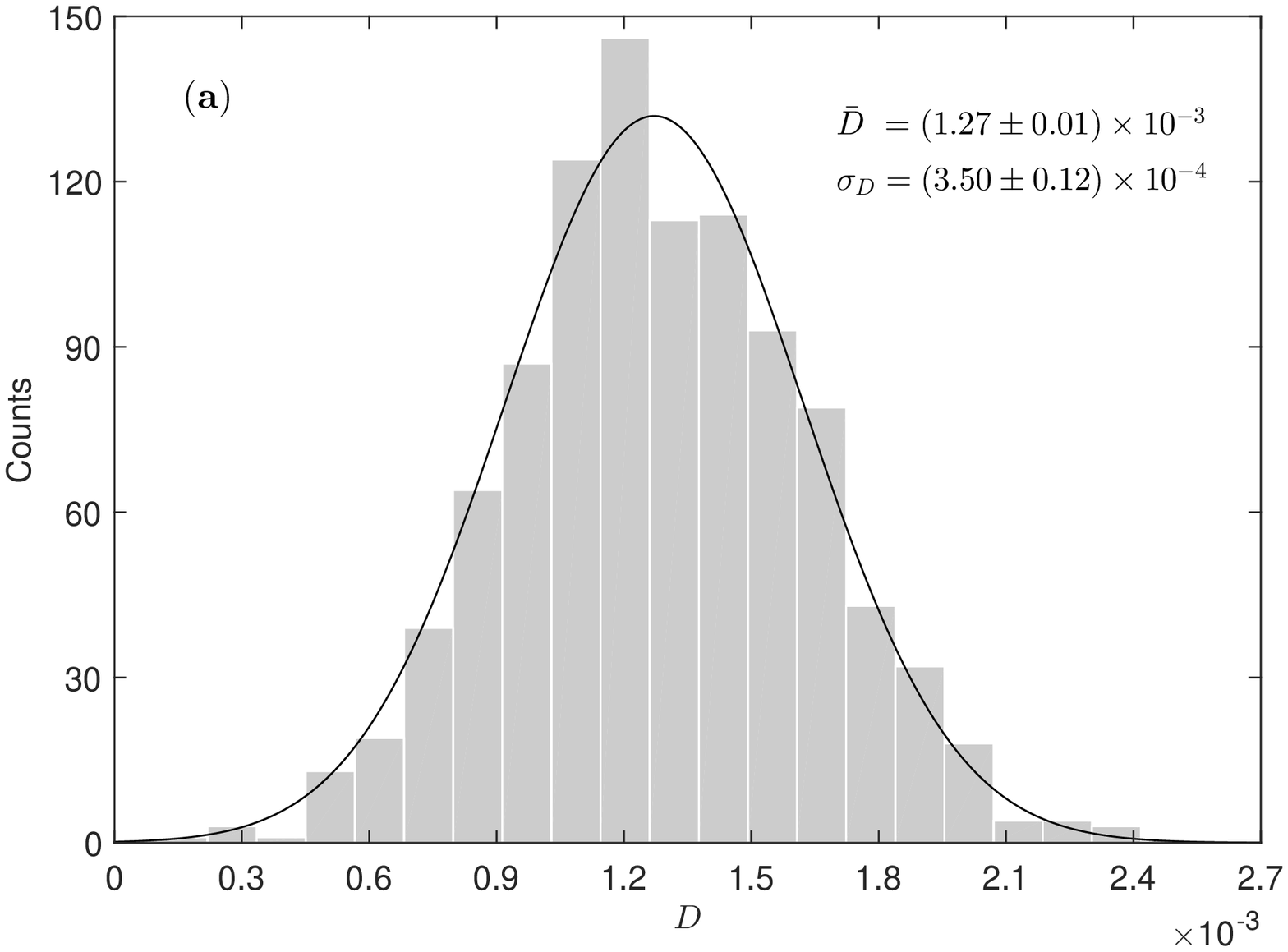}
  \includegraphics[width=0.5\textwidth]{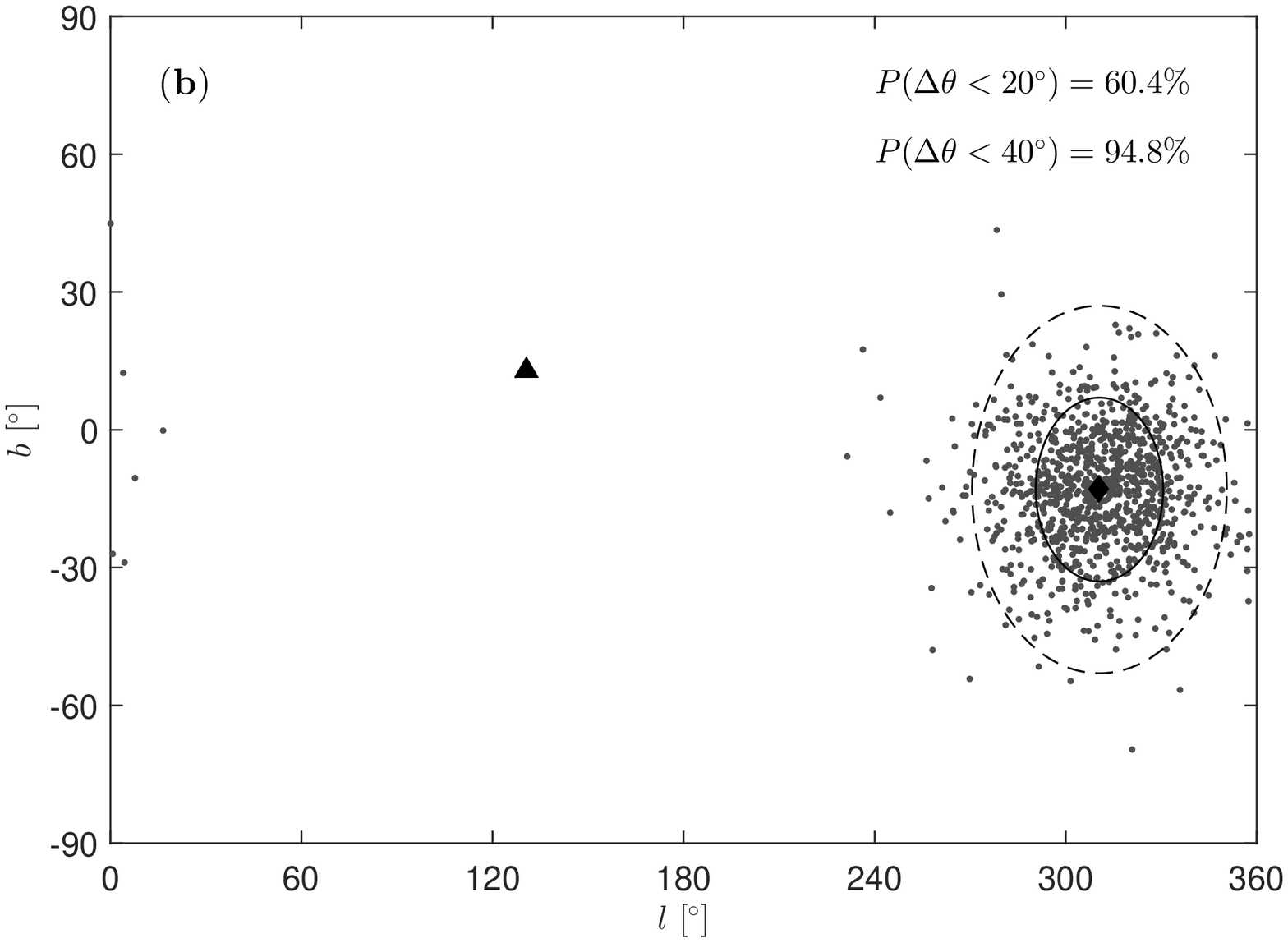}
  \caption{The dipoles of mock sample B in 1000 simulations. Panel (a): the histogram of dipole amplitudes, with black curve the best-fitting result to Gaussian function. Panel (b): the dipole directions in the sky of galactic coordinates. The black diamond is the fiducial dipole direction pointing towards $(l,b)=(310.6^{\circ},-13.0^{\circ})$, and the black triangle is the antipode. The solid (dashed) circle represents a circular region of radius $\Delta\theta<20^{\circ}$ ($\Delta\theta<40^{\circ}$), centering on the fiducial dipole direction.}
  \label{fig:DF2}
\end{figure}

\begin{figure}
  \centering
  \includegraphics[width=0.5\textwidth]{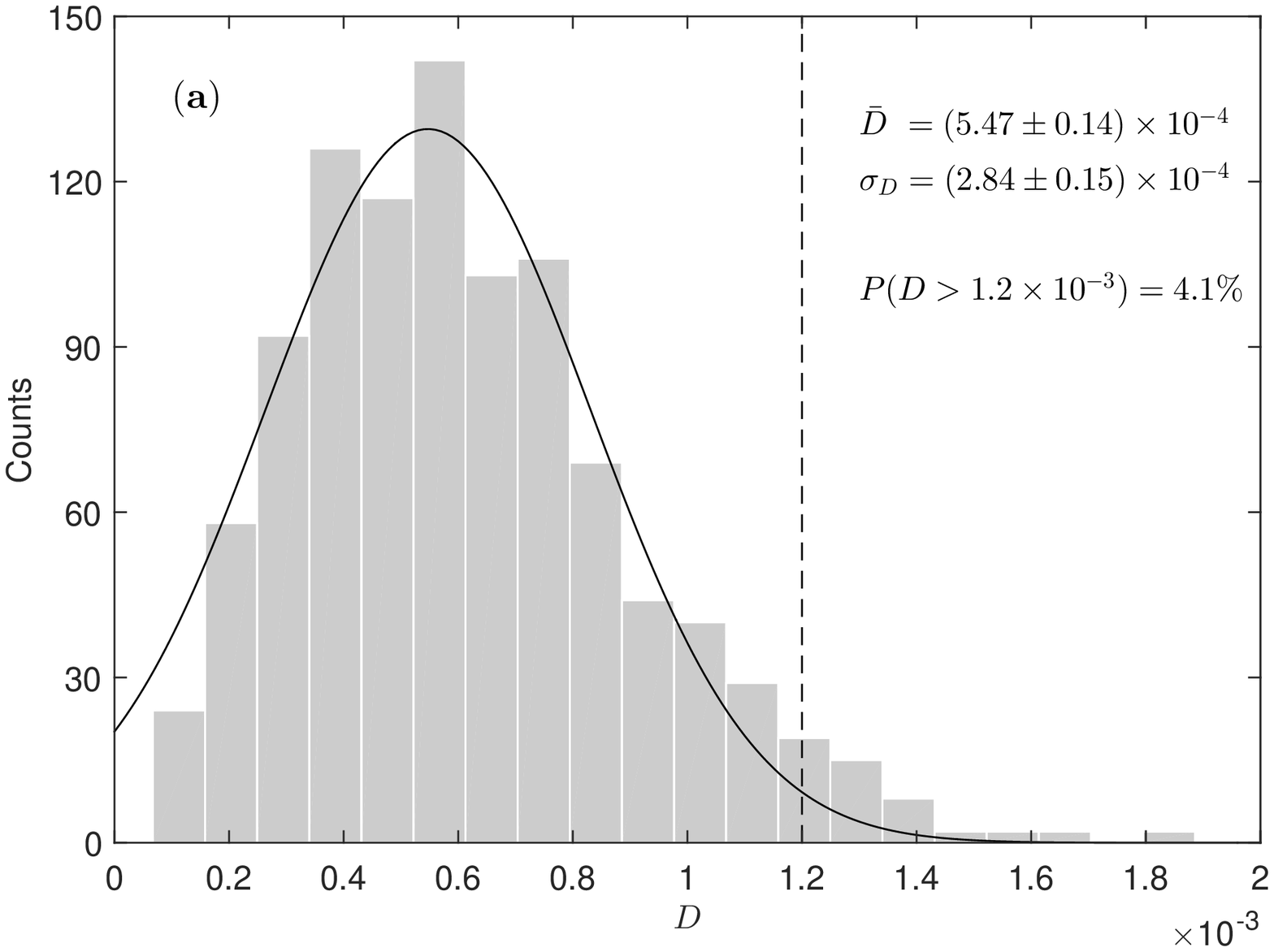}
  \includegraphics[width=0.5\textwidth]{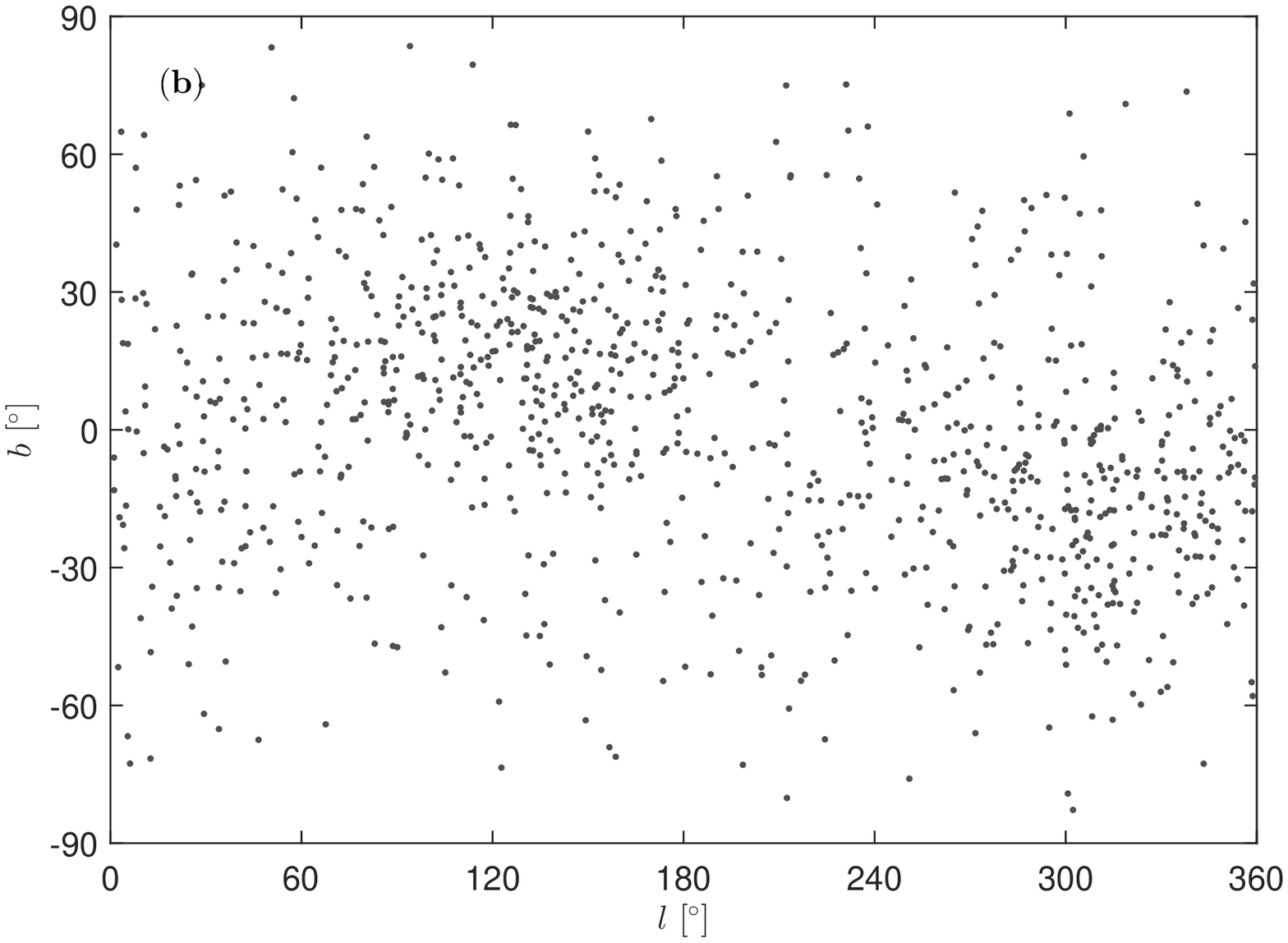}
  \caption{The dipoles of mock sample C in 1000 simulations. Panel (a): the histogram of dipole amplitudes, with black curve the best-fitting result to Gaussian function. The dashed vertical line represents the dipole amplitude of Union 2.1. Panel (b): the dipole directions in the sky of galactic coordinates.}
  \label{fig:DF3}
\end{figure}

\begin{figure}
  \centering
  \includegraphics[width=0.5\textwidth]{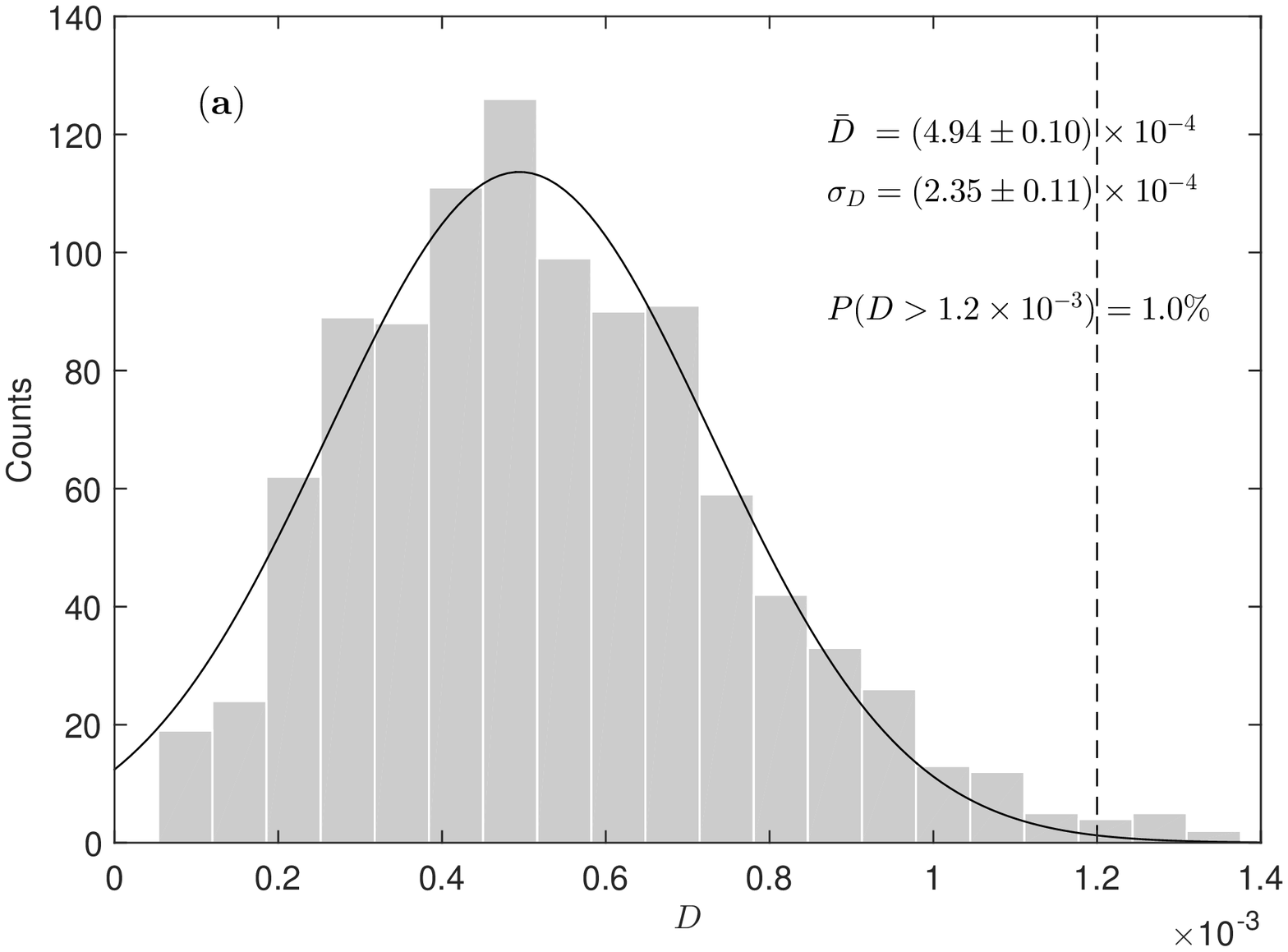}
  \includegraphics[width=0.5\textwidth]{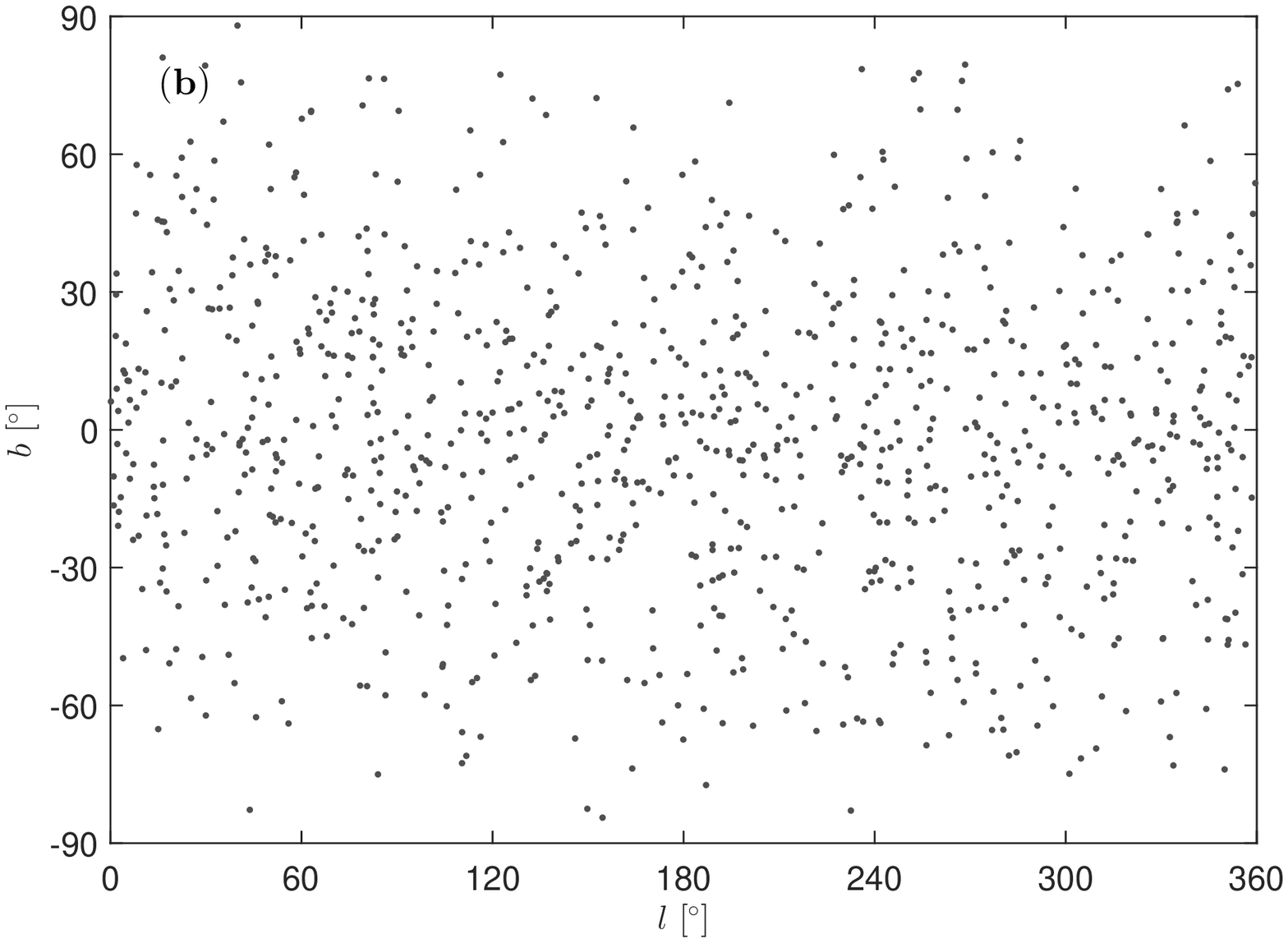}
  \caption{The dipoles of mock sample D in 1000 simulations. Panel (a): the histogram of dipole amplitudes, with black curve the best-fitting result to Gaussian function. The dashed vertical line represents the dipole amplitude of Union 2.1. Panel (b): the dipole directions in the sky of galactic coordinates.}
  \label{fig:DF4}
\end{figure}

Sample A is constructed to test whether the DF method can correctly reproduce the real dipole amplitude and direction when the data are indeed anisotropic. Sample C is constructed to test whether the DF method can detect pseudo anisotropic signals when the data are actually isotropic. Sample B and sample D are necessary to check if the inhomogeneous distribution of data points has some influences on the anisotropic signals.

Now we replace the Union2.1 data set with mock sample A, and search for the anisotropic signals using DF method as discussed above. The results in 1000 MC simulations are depicted in Fig. \ref{fig:DF1}. Panel (a) is the histogram of dipole amplitudes. It can be well fitted by the Gaussian function, with an average value $\bar{D}=(1.25\pm 0.02)\times 10^{-3}$ and standard deviation $\sigma_D=(5.07\pm 0.17)\times 10^{-4}$. The average dipole amplitude is well consistent with that of Union2.1. Panel (b) shows the mock dipole directions (black dots) in the sky of galactic coordinates. The black diamond is the fiducial dipole direction pointing towards $(l,b)=(310.6^{\circ},-13.0^{\circ})$, and the black triangle is the antipode. About $60.3\%$ realizations fall into a circular region of radius $\Delta\theta<20^{\circ}$ (the solid circle) centering on the fiducial dipole direction. If we enlarge the radius to $40^{\circ}$ (the dashed circle), the probability increases to $90.4\%$. In very small realizations, the mock dipole direction falls into the hemisphere completely opposite to the fiducial dipole direction (i.e, with $\Delta\theta>90^{\circ}$). This implies that DF method can correctly reproduce the dipole amplitude and direction.

Then we apply DF method to mock sample B. The results of 1000 MC simulations are plotted in Fig. \ref{fig:DF2}. The dipole amplitudes follow the Gaussian distribution, with an average value $\bar{D}=(1.27\pm 0.01)\times 10^{-3}$ and standard deviation $\sigma_D=(3.50\pm 0.12)\times 10^{-4}$, see panel (a). The value $\bar{D}$ of mock sample B is well consistent with that of mock sample A, while the value $\sigma_D$ is relatively smaller. From panel (b), we can clearly see that all the mock dipole directions cluster near the fiducial dipole direction. About $60.4\%$ ($94.8\%$) mock dipole directions fall into a circular region of radius $\Delta\theta<20^{\circ}$ ($\Delta\theta<40^{\circ}$). This probability is a little higher compared to that of mock sample A. Non of the mock dipole direction is more than $90^{\circ}$ away from the fiducial dipole direction. Therefore, DF method can well detect the anisotropic signals if the data points are homogeneously distributed in the sky.

Next, DF method is applied to mock sample C, and the results are depicted in Fig. \ref{fig:DF3}. Panel (a) is the histogram of dipole amplitudes in 1000 MC simulations. It can be fitted to the Gaussian function, with the best-fitting values $\bar{D}=(5.47\pm 0.14)\times 10^{-4}$, and $\sigma_D=(2.84\pm 0.15)\times 10^{-4}$. This means that, due to the statistical noise, DF method may detect pseudo anisotropic signals even if the data are actually isotropic. However, only $4.1\%$ realizations have dipole amplitudes larger than that of Union2.1. Hence, the probability that the anisotropic signal of Union2.1 is purely arising from statistical noise is very small. Panel (b) shows that the mock dipole directions are homogeneously distributed in the sky, as is expected.

Finally, we apply DF method to mock sample D, and plot the results of 1000 MC simulations in Fig. \ref{fig:DF4}. Similar to sample C, the dipole magnitudes of sample D also follow the Gaussian distribution, with an average value $\bar{D}=(4.94\pm 0.10)\times 10^{-4}$, and standard deviation $\sigma_D=(2.35\pm 0.11)\times 10^{-4}$, see panel (a). Even if the data points are homogeneously distributed in the sky, DF method may lead to false detection due to the statistical noise. However, the probability of such a false detection is as small as $1\%$. Panel (b) shows that the mock dipole directions are homogeneously distributed in the sky.

In summary, the sky distribution of supernovae in Union2.1 is extremely inhomogeneous. However, such an inhomogeneous distribution does not significantly bias the anisotropic signals obtained using DF method. The anisotropic signals hiding in Union2.1 couldn't be purely explained by the statistical noise.

\section{Hemisphere comparison}\label{sec:hemisphere}

This section is focus on discussing the HC method. We first shortly introduce the HC method. Then we use HC method to search for the anisotropic signals hiding in Union2.1 data set. Finally, we use four mock samples constructed in the last section to test the statistical significance of our results.

The main idea of HC method is to divide the sky into two opposite hemispheres, and fit the data points in each hemisphere to a specific cosmological model (e.g., $\Lambda$CDM model) separately. Then compare the cosmological parameters (such as $\Omega_M$) between different hemispheres and find the hemispheres with most discrepancy. The detailed procedures are as follows:
\begin{enumerate}
  \item{Given any direction $\hat{\bm n}$, the corresponding `equator' cuts the sky into two hemispheres without intersection, which we call `up' hemisphere and `down' hemisphere, respectively. Hence, the data points are divided into two subgroups according to their positions in the sky.}
  \item{Fit each subgroup to $\Lambda$CDM model separately, and derive the best-fitting parameters $\Omega_{M,u}$ and $\Omega_{M,d}$, where the subscripts `u' and `d' mean that the quantity is derived by fitting to the data points in the `up' hemisphere and `down' hemisphere, respectively.
      Define the anisotropic amplitude in this direction as
      \begin{equation}\label{eq:DOmegaM}
        D_{\Omega_M}(\hat{\bm n})\equiv\left|\frac{\Delta\Omega_M}{\bar{\Omega}_M}\right| =\left|\frac{\Omega_{M,u}-\Omega_{M,d}}{(\Omega_{M,u}+\Omega_{M,d})/2}\right|.
      \end{equation}}
  \item{Let $\hat{\bm n}$ runs over the whole sky, and find the direction which can maximize the value $D_{\Omega_M}$.}
\end{enumerate}

Here we give some notes. In the first step, when we use the name `equator', we mean the great circle perpendicular to the direction $\hat{\bm n}$. In the second step, when fitting the data to $\Lambda$CDM model, the Hubble parameter $H_0$ is marginalized using the method presented in \citet{CaiTuo:2012}. This is equivalent to free $H_0$, and fit $\Omega_M$ and $H_0$ simultaneously. Fixing $H_0$ does not significantly affect the results \citep{Chang:2015}. In the third step, due to the continuity of the sky, it is in practice impossible to let $\hat{\bm n}$ runs over the whole sky. One usual way is to divide the sky into $N_{\rm pix}$ pixels with equal area using the HEALPix method \citep{Gorski:2005}, and let $\hat{\bm n}$ runs over the center of each pixel. We choose $N_{\rm side}=8$ so that the sky is divided into $N_{\rm pix}=768$ pixels. This division has an angular resolution about $7.33^{\circ}$. Dividing the sky into more pixels can, but cannot significantly, improve the accuracy \citep{Chang:2015}.

We are in the way to investigate the anisotropy of Union2.1. We find the direction of maximum anisotropy points towards
\begin{equation}\label{eq:max_acceleration}
  (l,b)=(241.9^{\circ},-19.5^{\circ}), ~~\Omega_M=0.232,
\end{equation}
or equivalently, its antipode
\begin{equation}\label{eq:min_acceleration}
  (l,b)=(61.9^{\circ},19.5^{\circ}),~~\Omega_M=0.316.
\end{equation}
In this two directions, the anisotropic amplitude, as defined in equation (\ref{eq:DOmegaM}), is $D_{\Omega_M}=0.306$. In what follows, when we say the direction of maximum anisotropy, we always refer to the direction of smaller $\Omega_M$.

\begin{figure}
  \centering
  \includegraphics[width=0.5\textwidth]{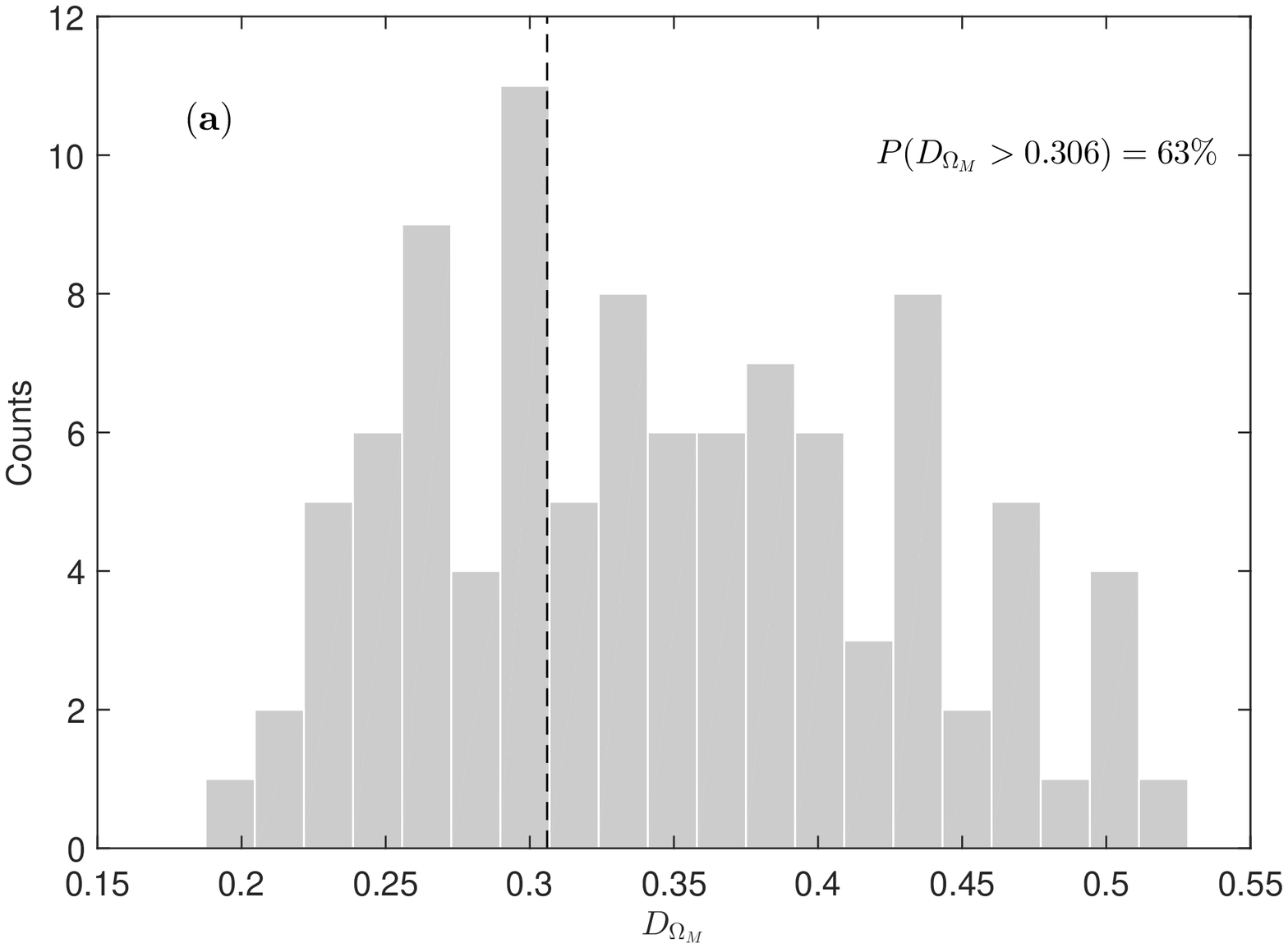}
  \includegraphics[width=0.5\textwidth]{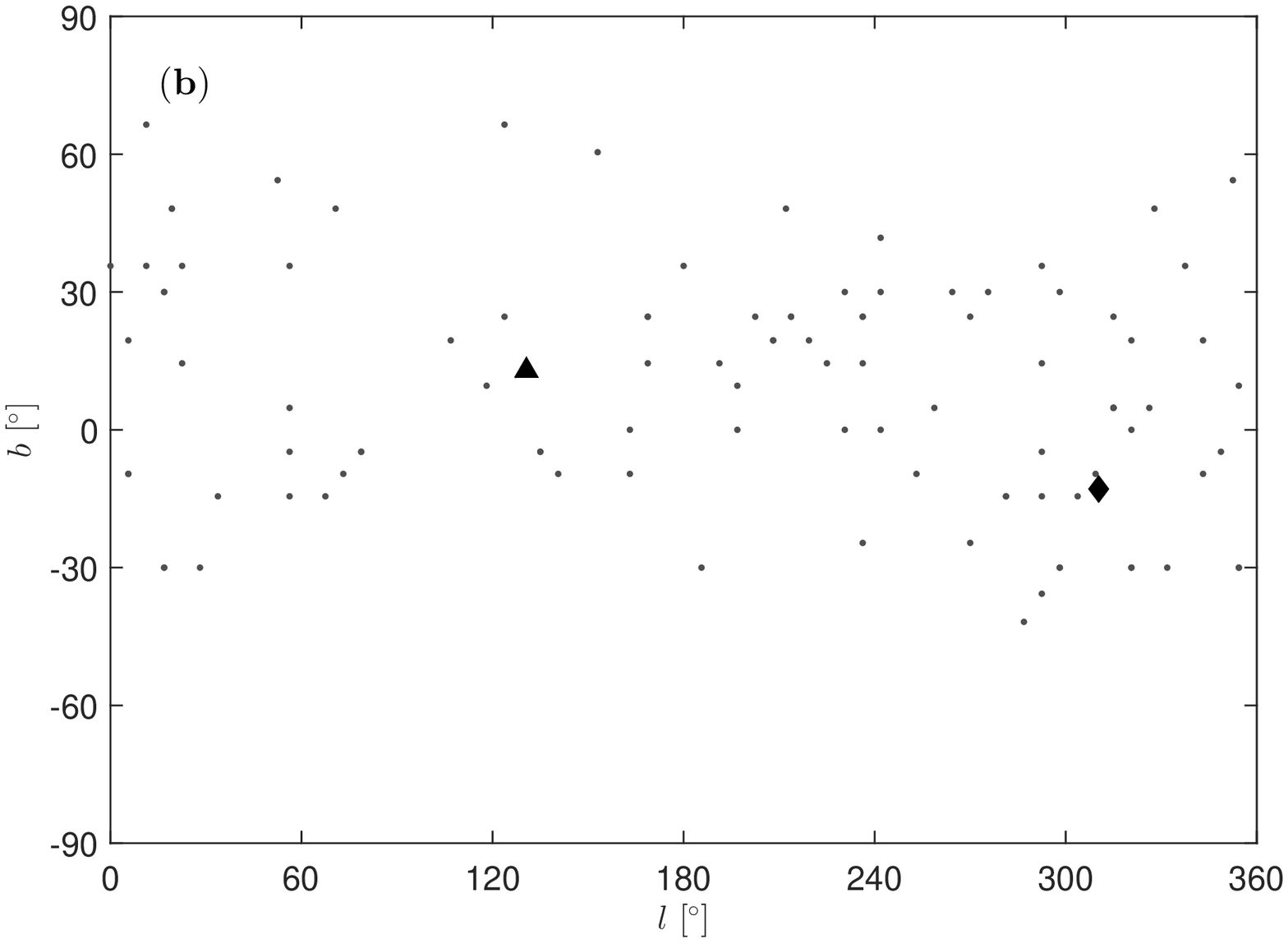}
  \caption{The anisotropic signal of mock sample A in 100 simulations, derived using the HC method. Panel (a): the histogram of anisotropic amplitudes. The dashed vertical line represents the anisotropic amplitude of Union 2.1. Panel (b): the directions of maximum anisotropy in the sky of galactic coordinates. The black diamond is the fiducial dipole direction pointing towards $(l,b)=(310.6^{\circ},-13.0^{\circ})$, and the black triangle is the antipode.}
  \label{fig:HC1}
\end{figure}

\begin{figure}
  \centering
  \includegraphics[width=0.5\textwidth]{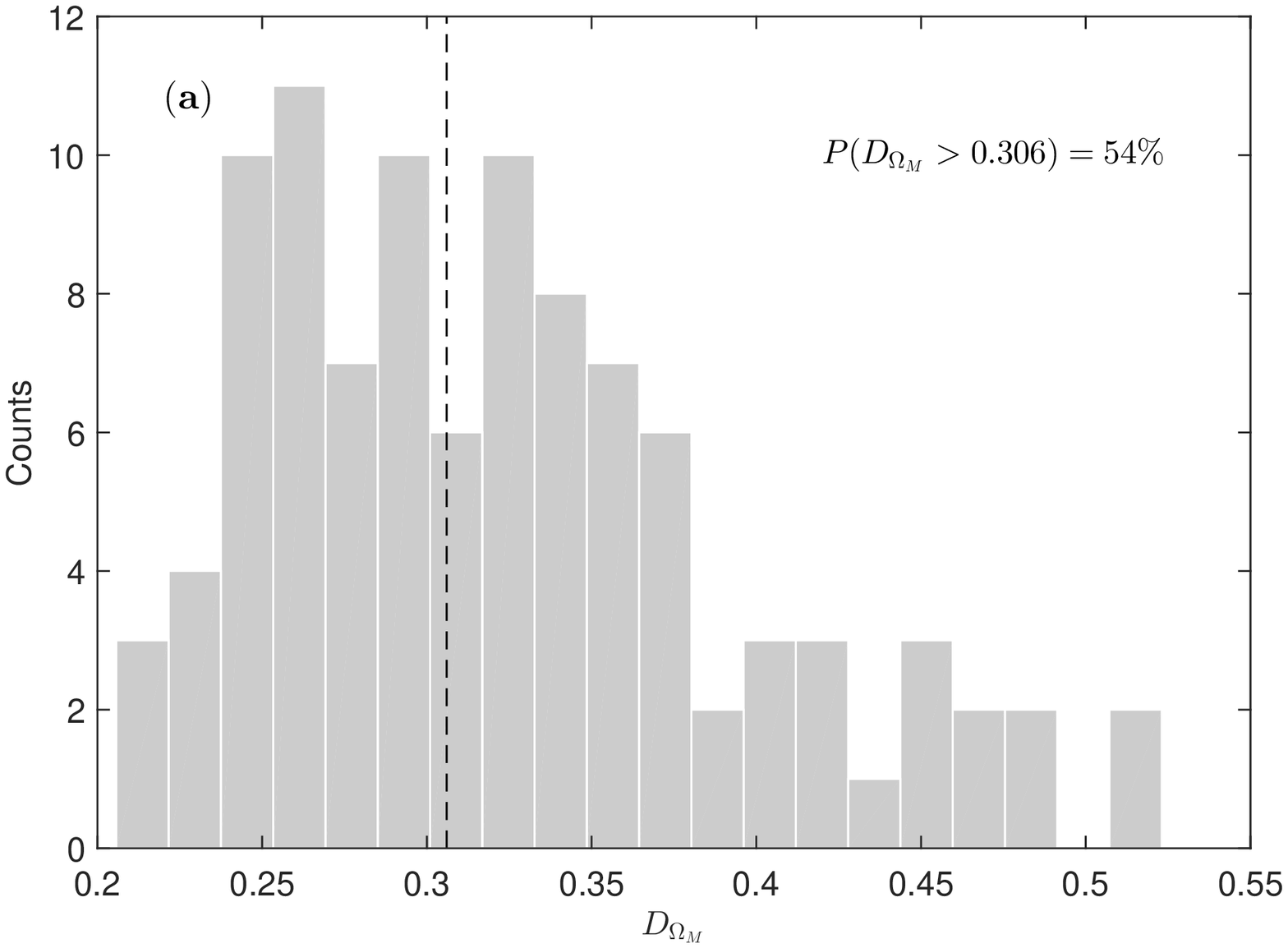}
  \includegraphics[width=0.5\textwidth]{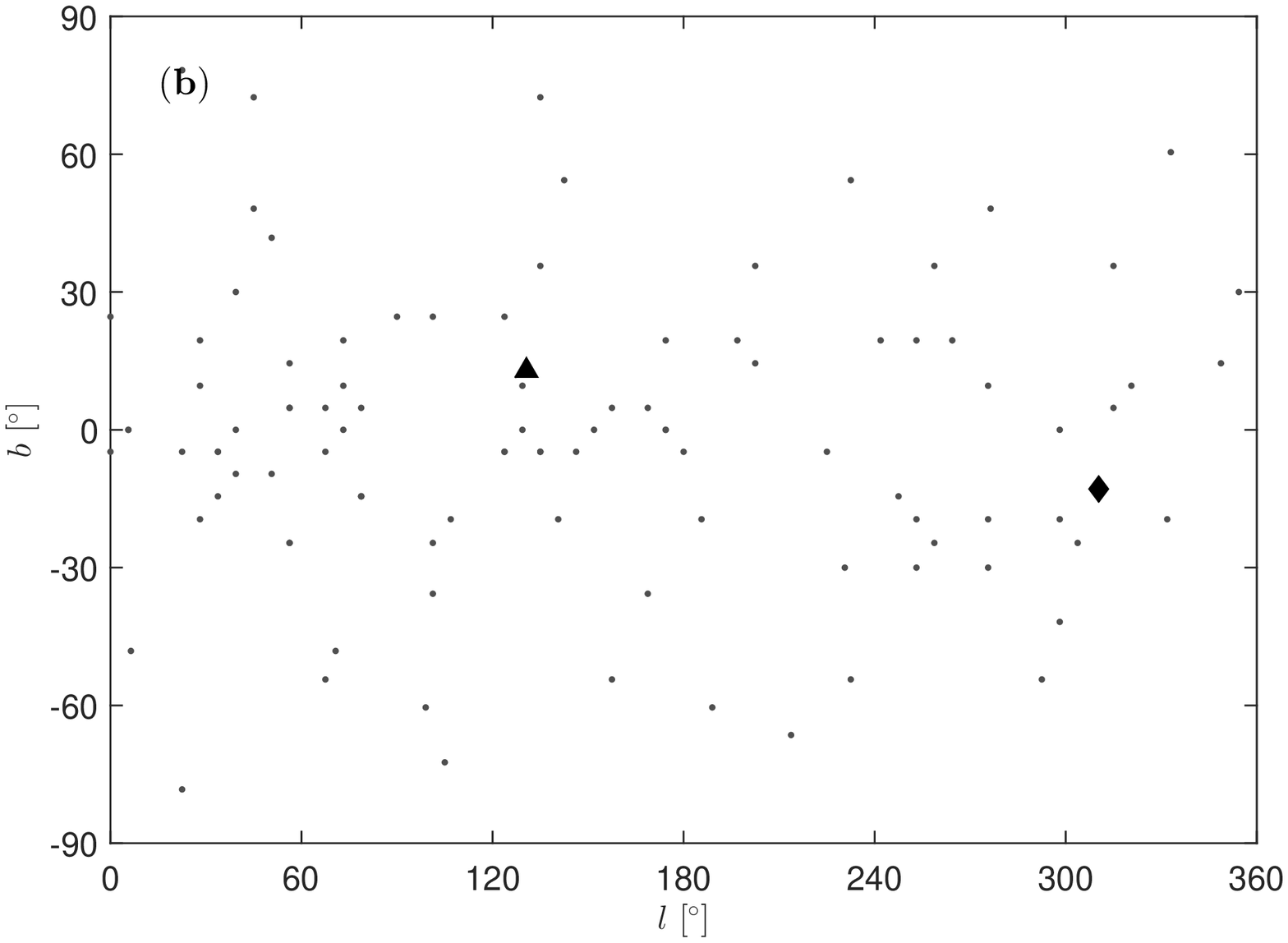}
  \caption{The anisotropic signal of mock sample B in 100 simulations, derived using the HC method. Panel (a): the histogram of anisotropic amplitudes. The dashed vertical line represents the anisotropic amplitude of Union 2.1. Panel (b): the directions of maximum anisotropy in the sky of galactic coordinates. The black diamond is the fiducial dipole direction pointing towards $(l,b)=(310.6^{\circ},-13.0^{\circ})$, and the black triangle is the antipode.}
  \label{fig:HC2}
\end{figure}

\begin{figure}
  \centering
  \includegraphics[width=0.5\textwidth]{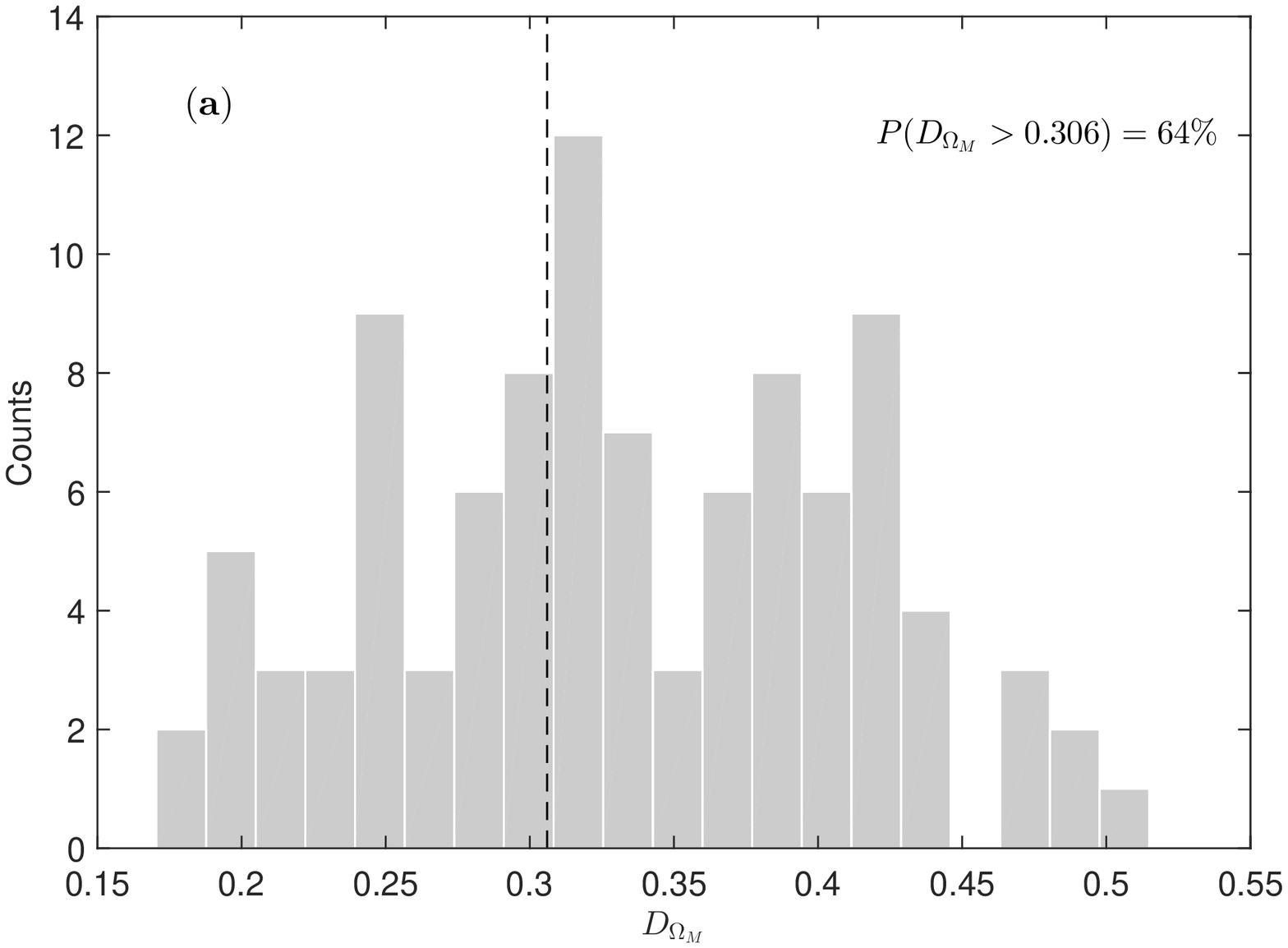}
  \includegraphics[width=0.5\textwidth]{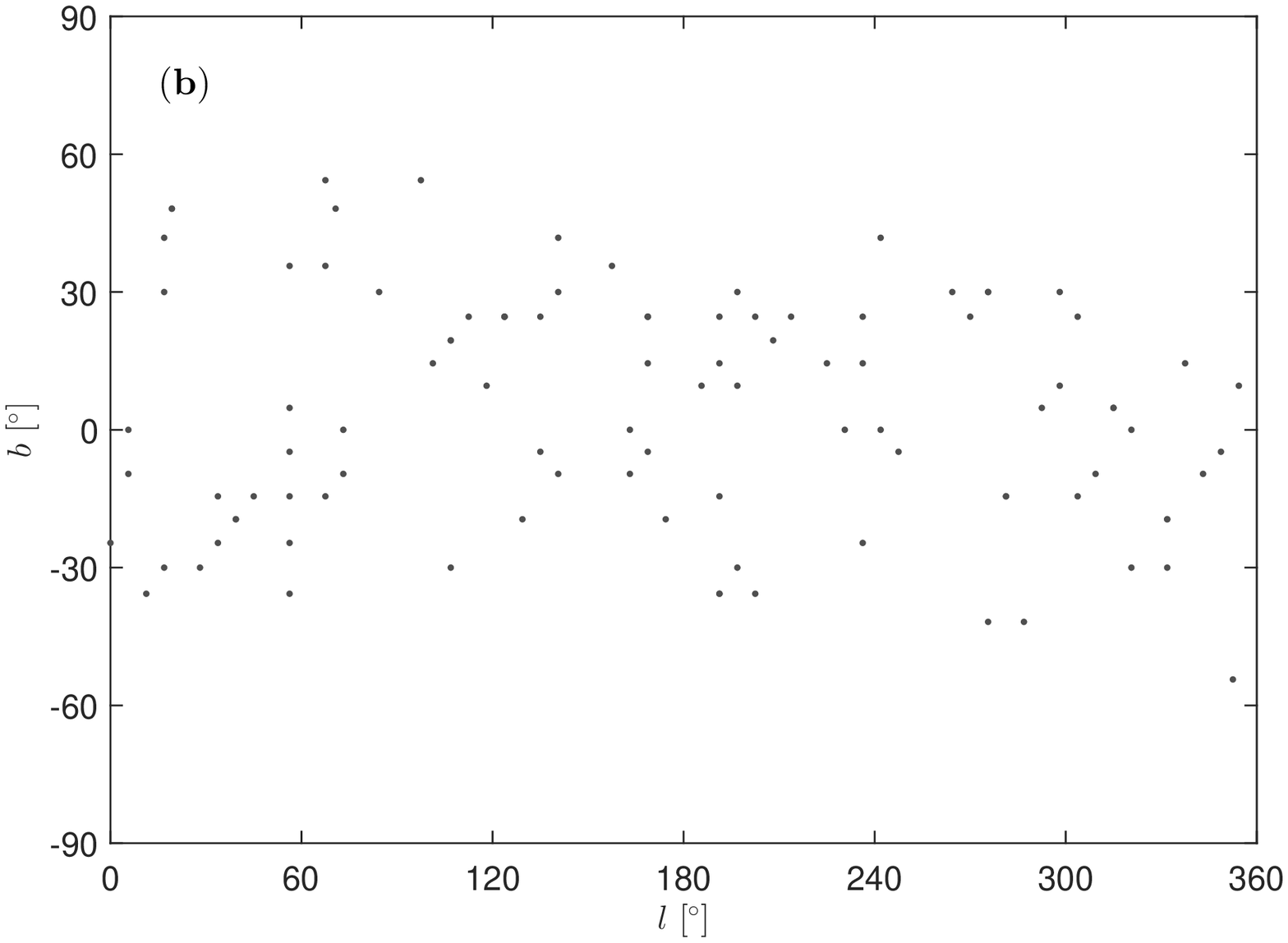}
  \caption{The anisotropic signal of mock sample C in 100 simulations, derived using the HC method. Panel (a): the histogram of anisotropic amplitudes. The dashed vertical line represents the anisotropic amplitude of Union 2.1. Panel (b): the directions of maximum anisotropy in the sky of galactic coordinates.}
  \label{fig:HC3}
\end{figure}

\begin{figure}
  \centering
  \includegraphics[width=0.5\textwidth]{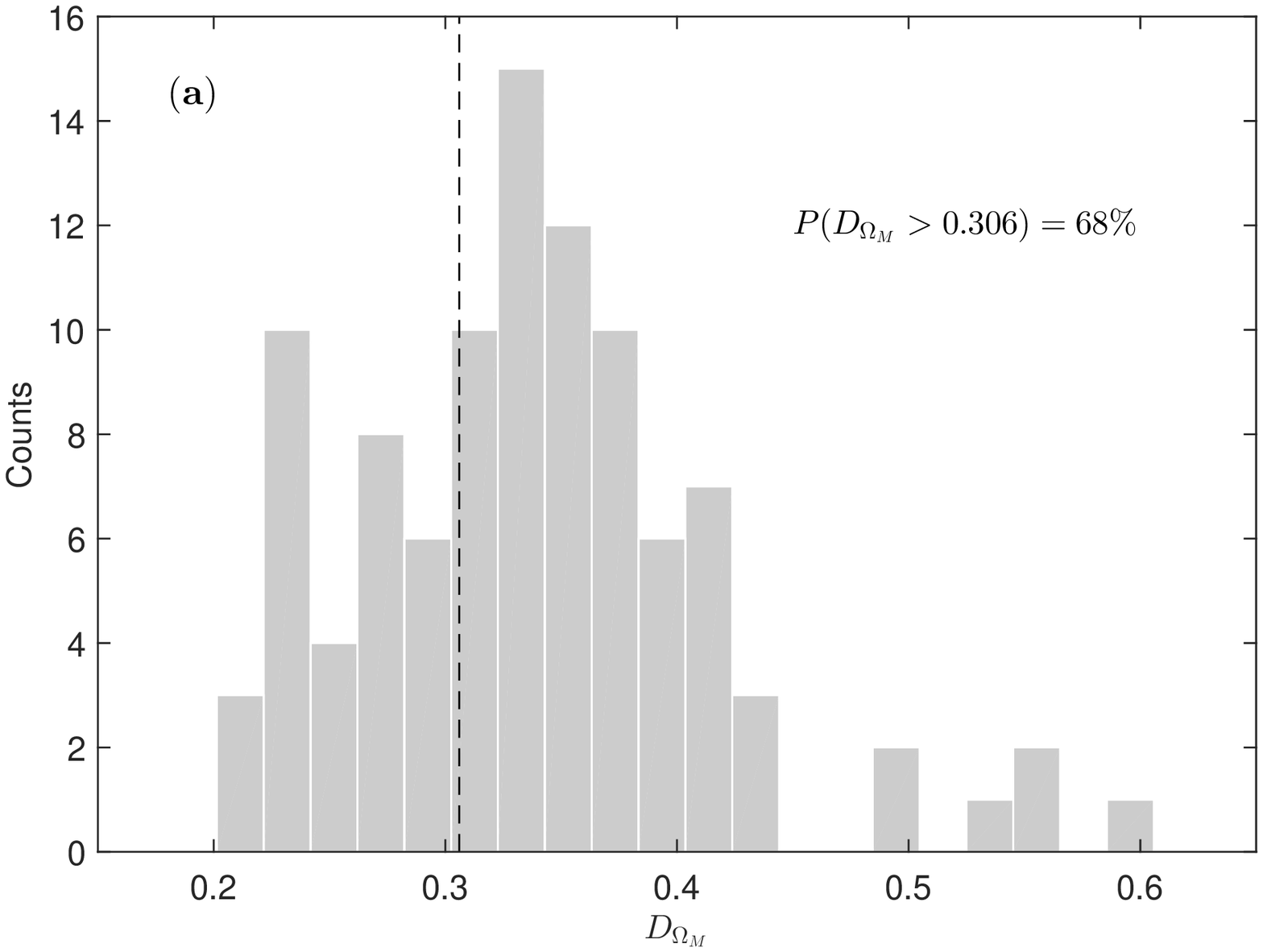}
  \includegraphics[width=0.5\textwidth]{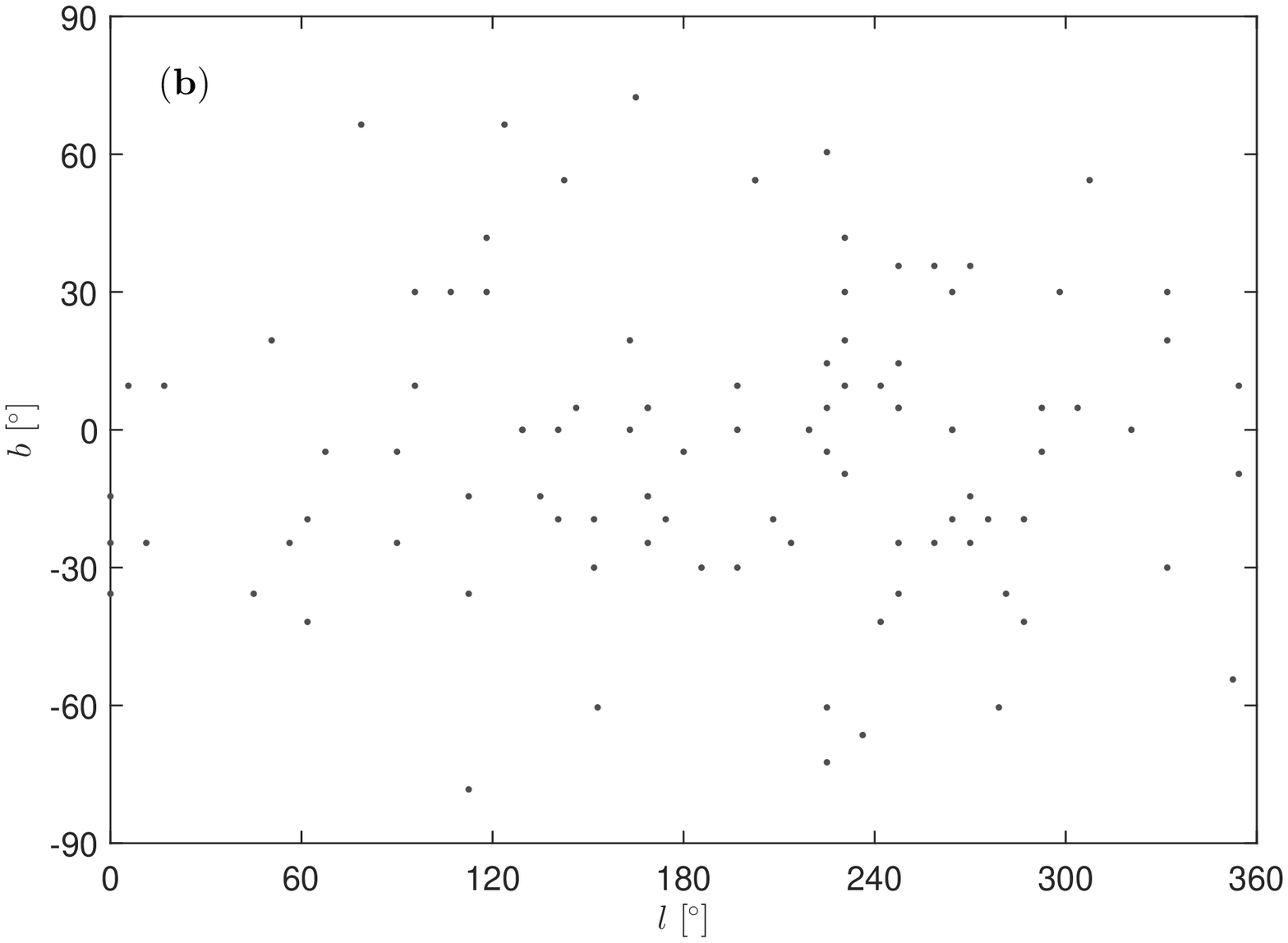}
  \caption{The anisotropic signal of mock sample D in 100 simulations, derived using the HC method. Panel (a): the histogram of anisotropic amplitudes. The dashed vertical line represents the anisotropic amplitude of Union 2.1. Panel (b): the directions of maximum anisotropy in the sky of galactic coordinates.}
  \label{fig:HC4}
\end{figure}

We note that the direction in equation (\ref{eq:max_acceleration}) is about $76^{\circ}$ away from that of Union2, while the later points to $(l,b)=(309^{\circ},18^{\circ})$ \citep{Antoniou:2010}. However, as is mentioned in the last section, the dipole direction of Union2.1 derived using DF method is well consistent with that of Union2. The Union2.1 data set is the updated version of Union2 by adding 23 more supernovae and by reducing the systematic uncertainty. The coincidence between both the dipole amplitudes and dipole directions of Union2.1 and Union2 strongly implies that the anisotropic signals hiding in these two data sets do not significantly differ from each other. Therefore, we suspect that the discrepancy between the directions of maximum anisotropy of Union2.1 and Union2 is due to the low statistical significance of HC method. To prove our hypothesis, we apply HC method to four mock samples.

First, we apply HC method to mock sample A. Note that HC method is much more computationally expansive than DF method. Due to the limited computational time, we just do 100 MC simulations. The results are plotted in Fig. \ref{fig:HC1}. Panel (a) is the distribution of anisotropic amplitudes, which span a wide range $0.188<D_{\Omega_M}<0.528$, and average at $\bar{D}_{\Omega_M}=0.346$. The average anisotropic amplitude, although a little larger, is still consistent with that of Union2.1. The probability that $D_{\Omega_M}$ is larger than $0.306$ is about $63\%$. The black dots in panel (b) are the directions of maximum anisotropy in 100 MC simulations. We also plot the fiducial dipole direction (black diamond) and its antipode (black triangle) for comparison. We can see that the directions of maximum anisotropy of mock samples are randomly distributed in the sky. This is beyond our expectation, because all the mock samples have the same fiducial dipole direction. This implies that HC method, although can approximately reproduce the anisotropic amplitude, in some cases may fail to reproduce the preferred direction.

The angular distribution of mock sample A is extremely inhomogeneous. To test if the homogeneous distribution of data points can improve this situation, we apply HC method to mock sample B. The results are plotted in Fig. \ref{fig:HC2}. The anisotropic amplitudes in 100 MC simulations are depicted in panel (a). The values of $D_{\Omega_M}$ fall into the range between $0.206$ and $0.523$, with an average value $0.323$. This is very close to the results of mock sample A, and consistent with the result of Union2.1. The probability that the mock amplitude is larger than $0.306$ is about $54\%$, which is a little smaller than that of mock sample A. Similar to mock sample A, the directions of maximum anisotropy of mock sample B are also randomly distributed in the sky, see panel (b). Therefore, we may come to the conclusion that the homogeneous distribution of data points couldn't significantly improve the reliability of HC method.

Next, we apply HC method to mock sample C. The results of 100 MC simulations are plotted in Fig. \ref{fig:HC3}. Panel (a) is the histogram of anisotropic amplitudes, and panel (b) is the directions of maximum anisotropy. The anisotropic amplitudes are in the range $0.171<D_{\Omega_M}<0.525$, with an average value $\bar{D}_{\Omega_M}=0.331$. Surprisingly, the average anisotropic amplitude of mock sample C is as large as that of sample A and sample B, and a little larger than that of Union2.1. The probability that $D_{\Omega_M}$ is larger than $0.306$ is about $64\%$, which is comparable to that of mock sample A and sample B. Since mock sample C is fiducially isotropic, the detected anisotropic signals must be caused by the statistical noise. As is expected, the directions of maximum anisotropy are randomly distributed in the sky.

Finally, HC method is used to search for the anisotropic signals of mock sample D. We plot the anisotropic amplitudes of 100 MC simulations in Fig. \ref{fig:HC4}(a). The average amplitude is $\bar{D}_{\Omega_M}=0.338$, which is as large as that of mock samples A, B and C. The probability that $D_{\Omega_M}$ is larger than $0.306$ is about $68\%$, the highest among four mock samples. This further suggests that the detected anisotropic signals in Union2.1, to a large extent, may be attributed to the statistical noise. Similar to the other three mock samples, the preferred directions of mock sample D are homogeneously distributed in the sky, see Fig. \ref{fig:HC4}(b).

In conclusion, HC method can approximately reproduce the anisotropic amplitude when the data points are really anisotropic. However, it may also detect the fake anisotropic signals when the data points are actually isotropic. Moreover, the direction of maximum anisotropy picked out by HC method is very likely deviating from the true direction. Therefore, the preferred direction found in Union2.1 using HC method is still suspicious.

\section{The dipole of mater density}\label{sec:matterdensity}

In the dipole direction of equation (\ref{dipole_direction}), the observed distance modulus is smaller than expected from $\Lambda$CDM model. In the opposite direction, however, it is larger than expected. This means that supernovae near the dipole direction are brighter than the ones near the opposite direction. Suppose this is induced by the anisotropic distribution of mater density, we may expect that matter near the dipole direction is denser than that in the opposite direction. However, the direction of the densest matter derived using HC method, given in equation (\ref{eq:min_acceleration}), is about $114^{\circ}$ away from the one derived using DF method. This phenomenon has already been noticed by \citet{Chang:2015} in the Union2 data set, where it was found that the directions of the brightest supernova obtained using DF method and HC method are approximately opposite.

To alleviate this discrepancy, we assume the distribution of matter is the dipole form, i.e.,
\begin{equation}\label{omegam_dipole}
  \Omega_M(\hat{\bm p})=\Omega_{M0}[1-A(\hat{\bm m}\cdot \hat{\bm p})],
\end{equation}
where $\Omega_{M0}$ is the average matter density, $A$ is the dipole amplitude, $\hat{\bm m}$ is the dipole direction, and $\hat{\bm p}$ is the position of supernova. To distinguish the dipole of matter density from the dipole of distance modulus which is discussed in section \ref{sec:dipole}, we call the former $\Omega_M$-dipole and the later $\mu$-dipole, respectively. If the $\mu$-dipole is mainly induced by the $\Omega_M$-dipole, we my expect that these two dipole directions are approximately opposite. The $\Omega_M$-dipole can arise from the Finslerian cosmology \citep{Li:2015}. Substituting equation (\ref{omegam_dipole}) into equations (\ref{eq:lumi_distance}) and (\ref{eq:distance_modulus}), we obtain the anisotropic distance modulus.

The best fit to Union2.1 data set leads to the $\Omega_M$-dipole amplitude
\begin{equation}\label{eq:OM_amplitude}
  A=0.160\pm 0.115,
\end{equation}
and the $\Omega_M$-dipole direction
\begin{equation}
  (l,b)=(171.8^{\circ}\pm 42.0^{\circ}, 9.9^{\circ}\pm 20.3^{\circ}).
\end{equation}
This result is obtained by fixing $\Omega_{M0}$ and $H_0$ to the values in equation (\ref{eq:parameter_iso}), and minimizing $\chi^2$ in equation (\ref{eq:chi2}). In this direction, $\Omega_M$ has the minimum
\begin{equation}\label{eq:OM_min}
  \Omega_{M,{\rm min}}=0.234\pm 0.032.
\end{equation}
The matter density has maximum at the opposite direction, i.e.,
\begin{equation}\label{eq:OM_max}
  (l,b)=(351.8^{\circ}\pm 42.0^{\circ}, -9.9^{\circ}\pm 20.3^{\circ}),
\end{equation}
with the largest matter density
\begin{equation}
  \Omega_{M,{\rm max}}=0.323\pm 0.032.
\end{equation}
With the notation of equation(\ref{eq:DOmegaM}), we obtain $D_{\Omega_M}=0.320\pm 0.164$, well consistent with that derived using HC method. However, the direction of minimum $\Omega_M$ we obtained here is about $75^{\circ}$ away from that obtained using HC method. The angle between the $\mu$-dipole direction and the antipole of the $\Omega_M$-dipole direction is about $40^{\circ}$, well consistent within $1\sigma$ uncertainty.

\begin{figure}
  \centering
  \includegraphics[width=0.5\textwidth]{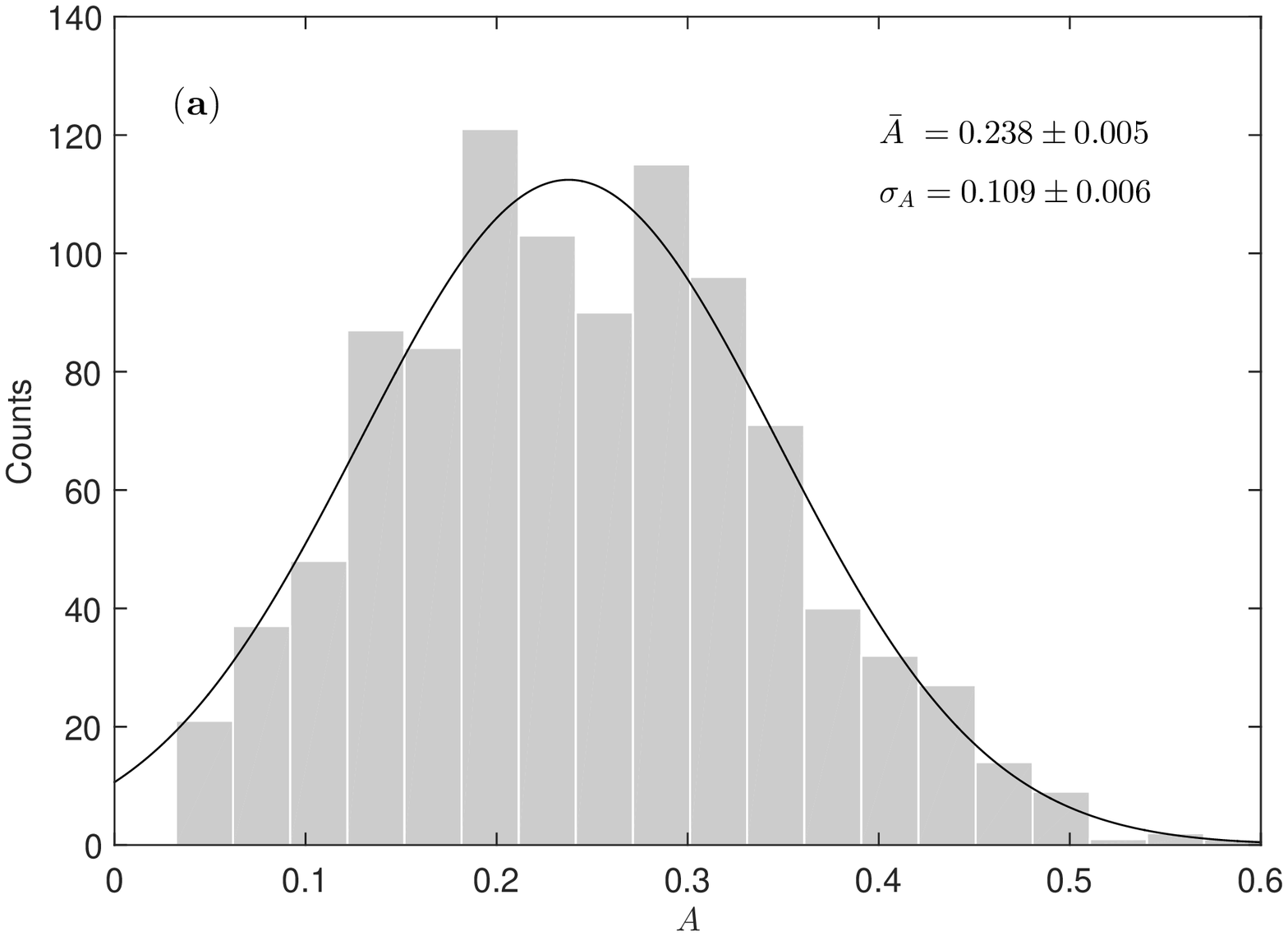}
  \includegraphics[width=0.5\textwidth]{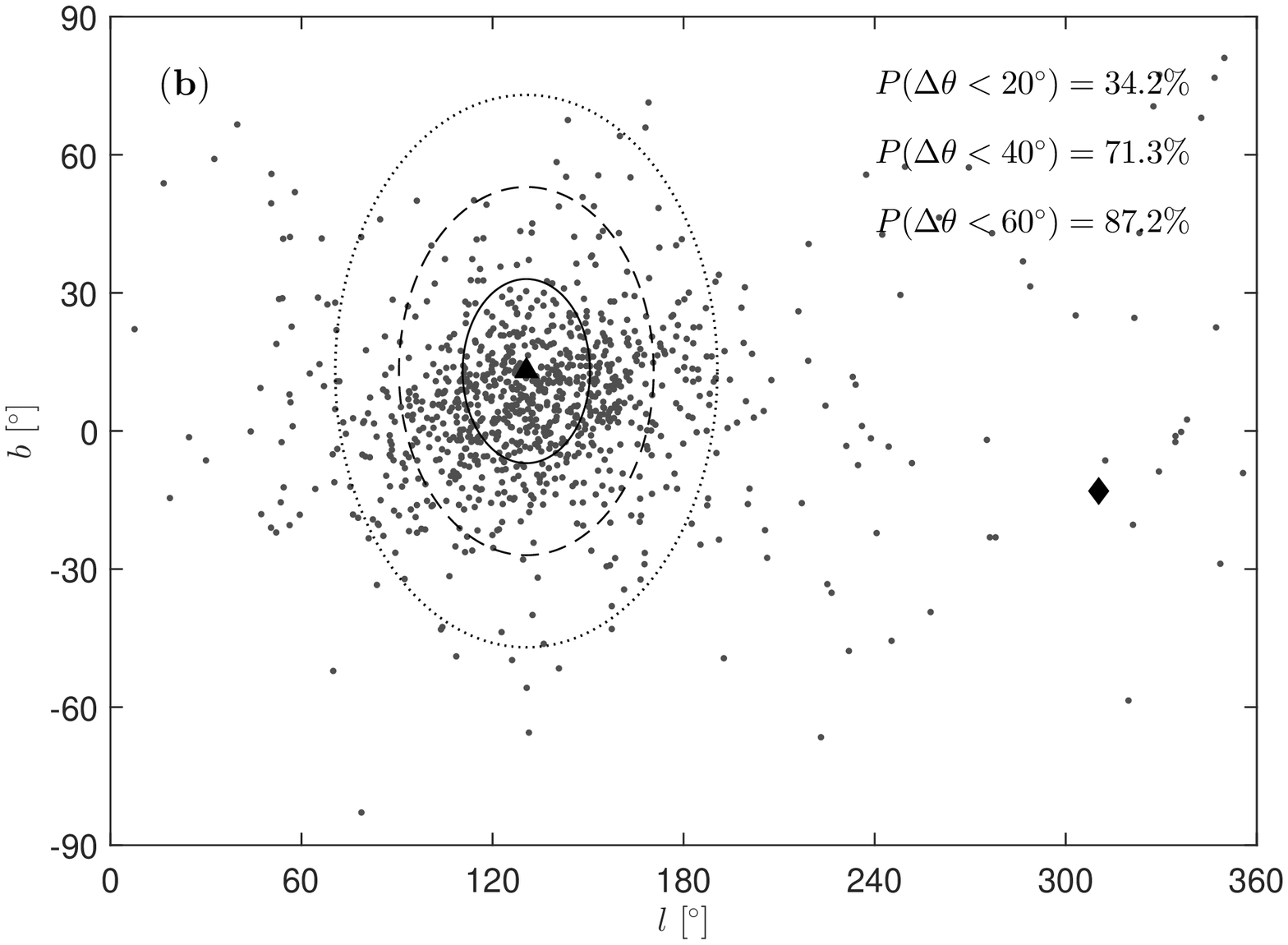}
  \caption{The $\Omega_M$-dipoles of mock sample A in 1000 simulations. Panel (a): the histogram of $\Omega_M$-dipole amplitudes, with black curve the best-fitting result to Gaussian function. Panel (b): the $\Omega_M$-dipole directions in the sky of galactic coordinates. The black diamond is the fiducial $\mu$-dipole direction pointing towards $(l,b)=(310.6^{\circ},-13.0^{\circ})$, and the black triangle is the antipode. The solid (dashed or dotted) circle represents a circular region of radius $\Delta\theta<20^{\circ}$ ($\Delta\theta<40^{\circ}$ or $\Delta\theta<60^{\circ}$), centering on the antipode of the fiducial $\mu$-dipole direction.}
  \label{fig:OM1}
\end{figure}

\begin{figure}
  \centering
  \includegraphics[width=0.5\textwidth]{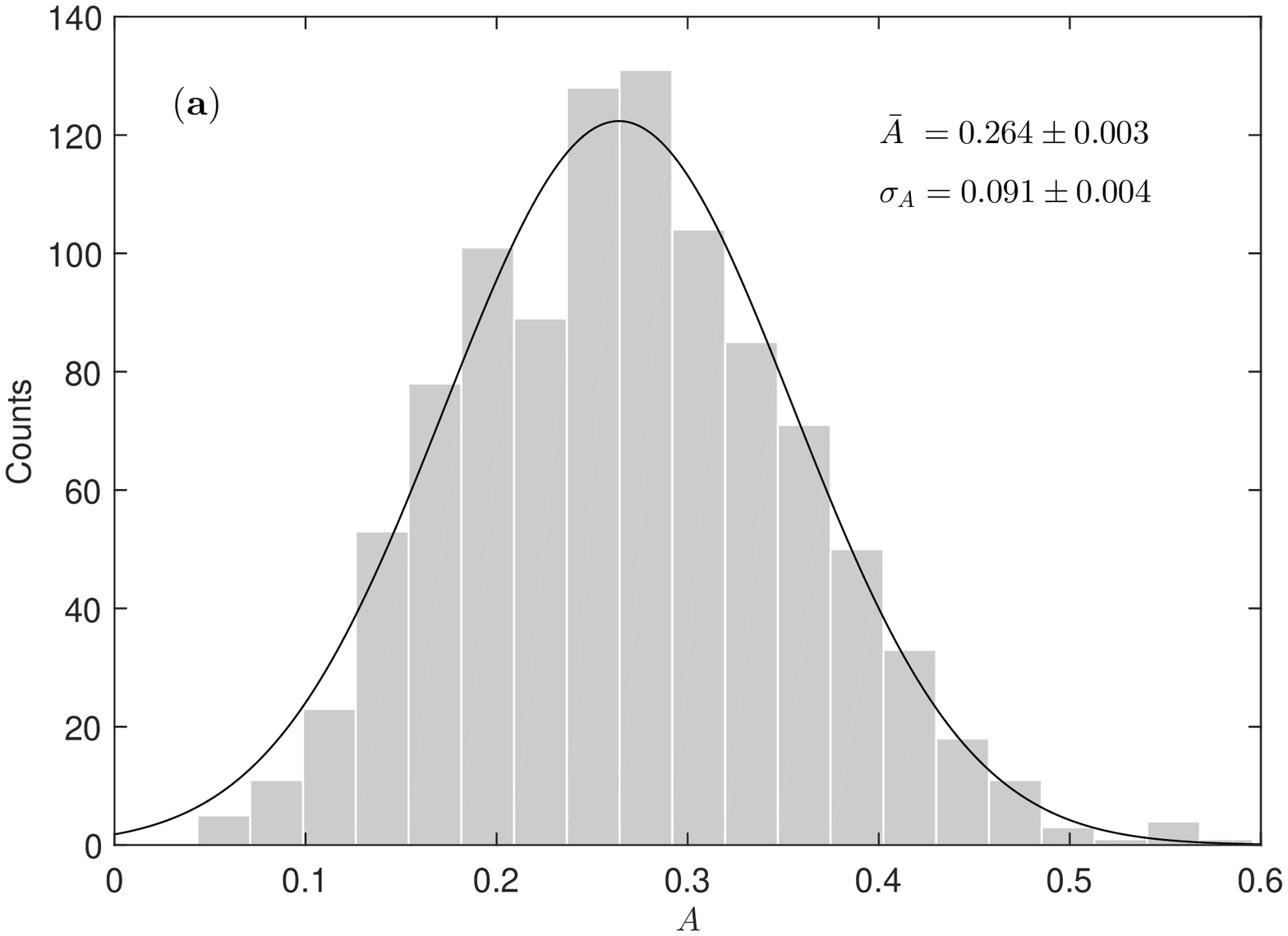}
  \includegraphics[width=0.5\textwidth]{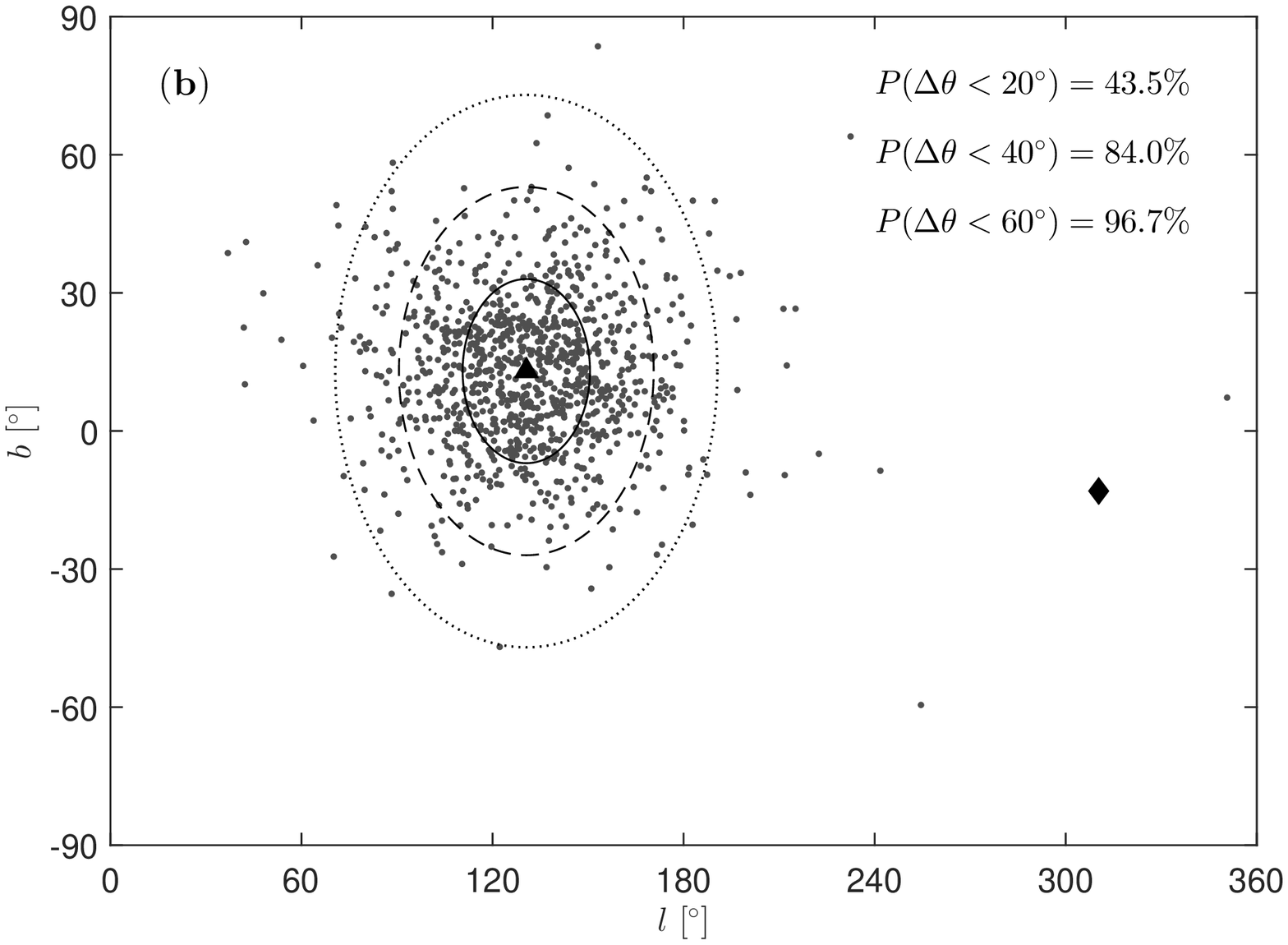}
  \caption{The $\Omega_M$-dipoles of mock sample B in 1000 simulations. Panel (a): the histogram of $\Omega_M$-dipole amplitudes, with black curve the best-fitting result to Gaussian function. Panel (b): the $\Omega_M$-dipole directions in the sky of galactic coordinates. The black diamond is the fiducial $\mu$-dipole direction pointing towards $(l,b)=(310.6^{\circ},-13.0^{\circ})$, and the black triangle is the antipode. The solid (dashed or dotted) circle represents a circular region of radius $\Delta\theta<20^{\circ}$ ($\Delta\theta<40^{\circ}$ or $\Delta\theta<60^{\circ}$), centering on the antipode of the fiducial $\mu$-dipole direction.}
  \label{fig:OM2}
\end{figure}

\begin{figure}
  \centering
  \includegraphics[width=0.5\textwidth]{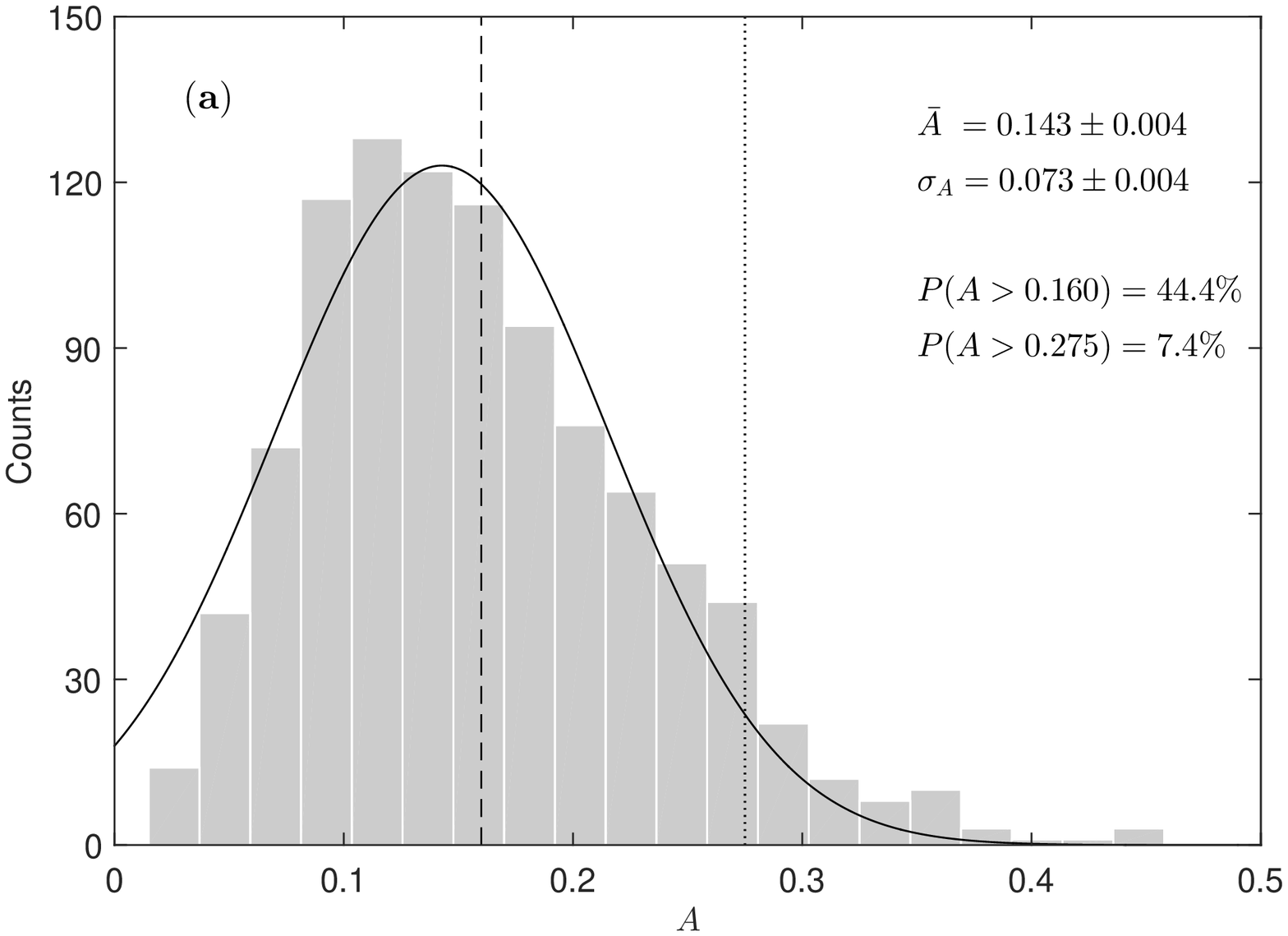}
  \includegraphics[width=0.5\textwidth]{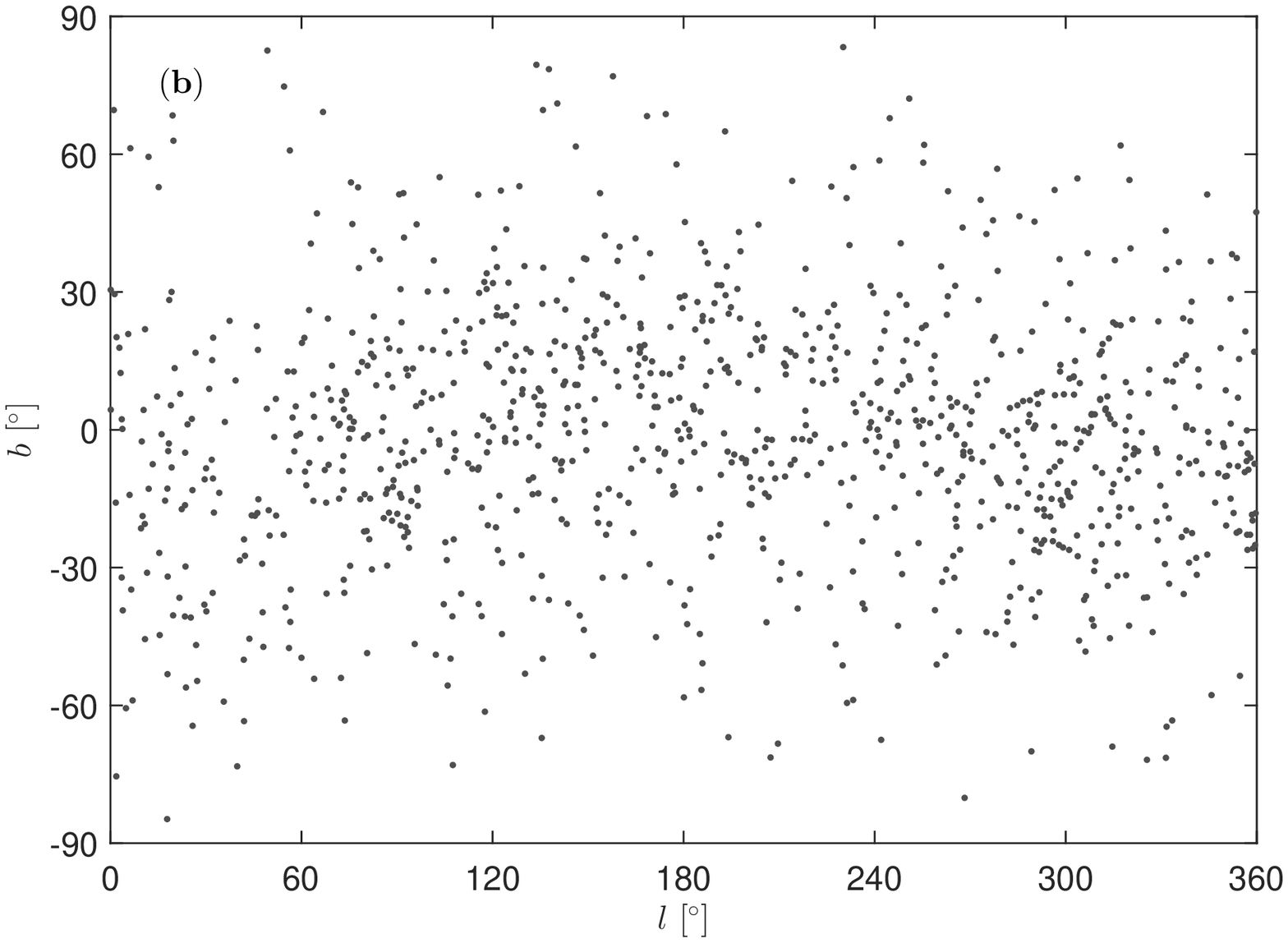}
  \caption{The $\Omega_M$-dipoles of mock sample C in 1000 simulations. Panel (a): the histogram of $\Omega_M$-dipole amplitudes, with black curve the best-fitting result to Gaussian function. The dashed (dotted) vertical line represents the central value (1$\sigma$ upper limit) of $\Omega_M$-dipole amplitude of Union 2.1. Panel (b): the $\Omega_M$-dipole directions in the sky of galactic coordinates.}
  \label{fig:OM3}
\end{figure}

\begin{figure}
  \centering
  \includegraphics[width=0.5\textwidth]{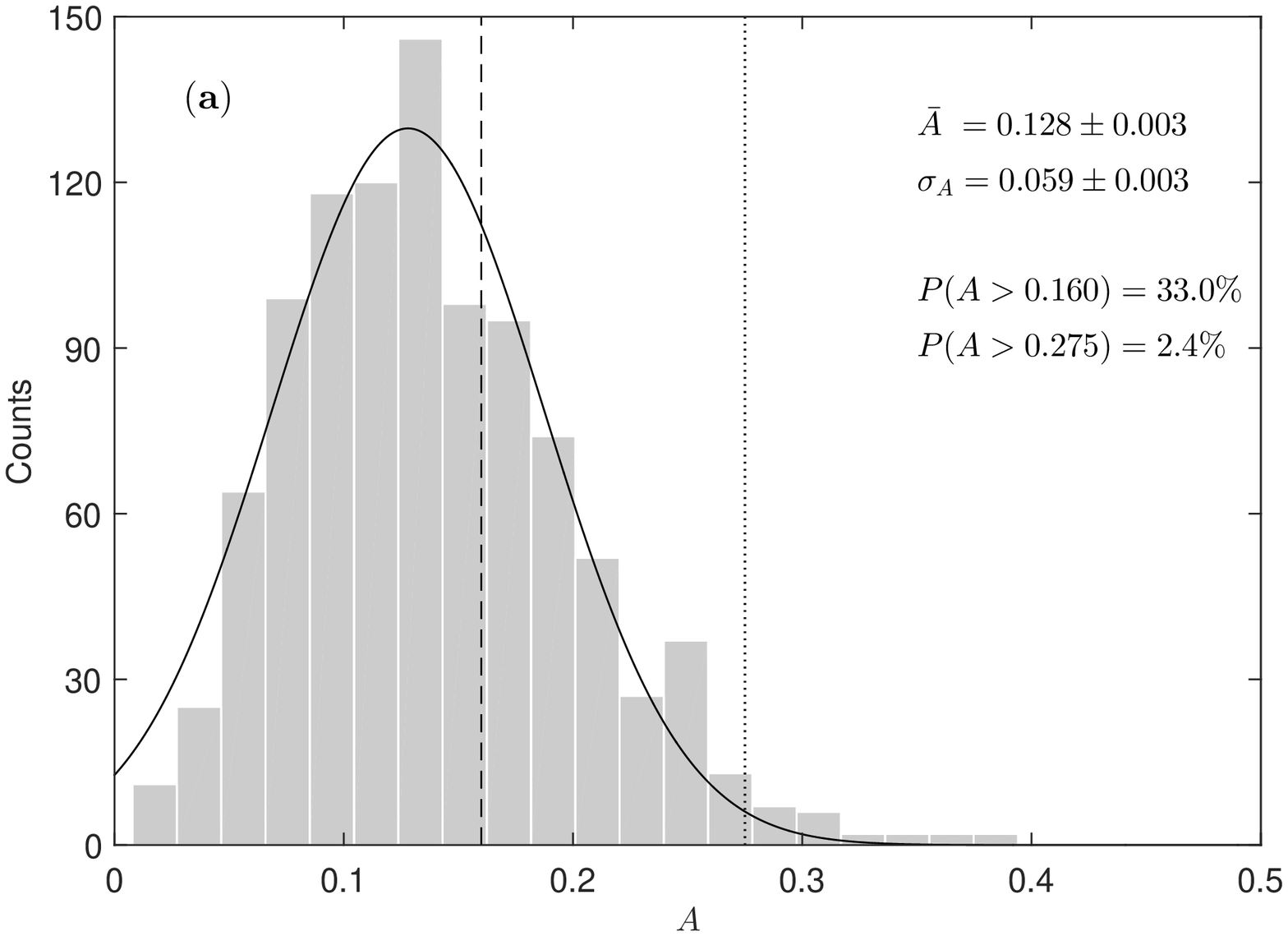}
  \includegraphics[width=0.5\textwidth]{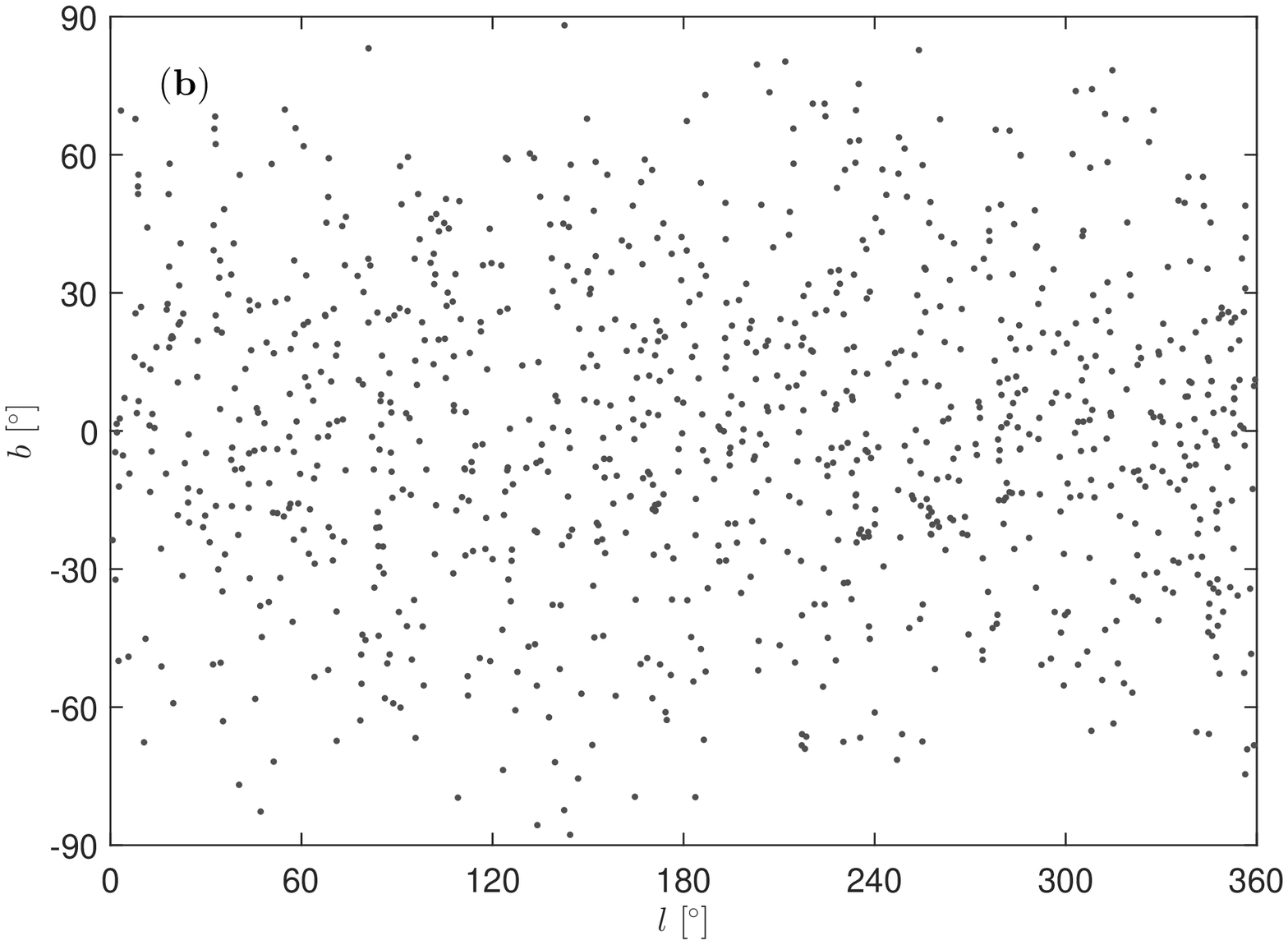}
  \caption{The $\Omega_M$-dipoles of mock sample D in 1000 simulations. Panel (a): the histogram of $\Omega_M$-dipole amplitudes, with black curve the best-fitting result to Gaussian function. The dashed (dotted) vertical line represents the central value (1$\sigma$ upper limit) of $\Omega_M$-dipole amplitude of Union 2.1. Panel (b): the $\Omega_M$-dipole directions in the sky of galactic coordinates.}
  \label{fig:OM4}
\end{figure}

Four mock samples are used to test the statistical significance of our results. First, we search for the $\Omega_M$-dipole in mock sample A. The results of 1000 MC simulations are plotted in Fig. \ref{fig:OM1}. Panel (a) is the histogram of $\Omega_M$-dipole amplitudes, which approximately follows the Gaussian distribution, with an average value $\bar{A}=0.238\pm 0.005$ and standard deviation $\sigma_A=0.109\pm 0.006$. Although $\bar{A}$ is a little larger than the $\Omega_M$-dipole amplitude of Union2.1, they are still consistent within $1\sigma$ uncertainty. The black dots in panel (b) show the $\Omega_M$-dipole directions in the sky. As is expected, they cluster near the antipode of the $\mu$-dipole direction (black triangle). The probabilities that the angle (denoted by $\Delta\theta$) between the mock direction and the fiducial direction is smaller than $20^{\circ}$, $40^{\circ}$, $60^{\circ}$ are $34.2\%$, $71.3\%$, $87.2\%$, respectively. There is only $5\%$ probability that $\Delta\theta$ is larger than $90^{\circ}$. This implies that the statistical significance of our method, although is relatively lower than DF method, is much higher than HC method.

To test if the homogeneous distribution of data points can improve our results, we search for the $\Omega_M$-dipole in mock sample B. The amplitudes and directions of $\Omega_M$-dipole in 1000 MC simulations are plotted in panel (a) and panel (b) of Fig. \ref{fig:OM2}, respectively. The amplitudes follow the Gaussian distribution, with an average values $\bar{A}=0.264\pm 0.003$, and standard deviation $\sigma_A=0.091\pm 0.004$. These are comparable to that of mock sample A. However, the mock directions are much more clustered than in the sample A case. The probabilities that $\Delta\theta$ is smaller than $20^{\circ}$, $40^{\circ}$, $60^{\circ}$ are $43.5\%$, $84.0\%$, $96.7\%$, respectively. There is a negligible probability ($\sim 0.4\%$) that $\Delta\theta>90^{\circ}$. This means that the statistical significance can be highly improved if the data points are homogeneously distributed.

Next, the same procedure is applied to mock sample C. The results of 1000 MC simulations are plotted in Fig. \ref{fig:OM3}. Panel (a) is the histogram of the $\Omega_M$-dipole amplitudes. It follows the Gaussian distribution with an average value $\bar{A}=0.143\pm 0.004$, and standard deviation $\sigma_A=0.073\pm 0.004$. The average amplitude of mock sample C is in agreement with that of Union2.1. The probability that the mock amplitude is larger than the amplitude of Union2.1 is $44.4\%$. Hence, similar to HC method, $\Omega_M$-dipole fitting may also lead to pseudo anisotropic signals when the data are actually isotropic. However, the $\Omega_M$-dipole of Union2.1 shows large uncertainty, with $1\sigma$ upper limit 0.275. Taking this into consideration, the probability that the mock amplitude is larger than 0.275 is only $7.4\%$. As is expected, the mock directions are almost homogeneously distributed in the sky, see panel (b) of Fig. \ref{fig:OM3}.

Finally, we apply our method to mock sample D. The results of 1000 MC simulations are depicted in Fig. \ref{fig:OM4}. Similar to mock sample C case, the amplitudes of mock sample D also follow the Gaussian distribution. The average value $\bar{A}=0.128\pm 0.003$ and standard deviation $\sigma_A=0.059\pm 0.003$ are a little smaller than that of mock sample C. The probabilities that $A>0.160$ and $A>0.275$ are $33.0\%$ and $2.4\%$, respectively. They are much smaller compared to mock sample C case. This further implies that the homogeneous distribution of data points can highly improve the statistical significance. Panel (b) is the distribution of mock directions in the sky. They seem to be more homogeneous than in the mock sample C case.

In conclusion, we obtained the amplitude and direction of $\Omega_M$-dipole in Union2.1 dat set. The direction of maximum $\Omega_M$ is aligned with the $\mu$-dipole direction, but it is more than $1\sigma$ away from the direction of maximum $\Omega_M$ resulting from HC method. MC simulation show that our method is much more statistically significant than HC method, although it is not as significant as DF method. If the data points are homogeneously distributed in the sky, the statistical significance can be further improved.

\section{Discussions and conclusions}\label{sec:discussion}

In this paper, we applied two different methods, i.e., DF and HC methods, to probe the anisotropic signals hiding in Union2.1 data set. We found that the directions of maximum $\Omega_M$ derived using DF and HC methods are $(l,b)=(310.6^{\circ}\pm 18.2^{\circ}, -13.0^{\circ}\pm 11.1^{\circ})$ and $(l,b)=(61.9^{\circ},19.5^{\circ})$, respectively. These two directions are about $114^{\circ}$ away from each other. Four Union2.1-like mock samples were constructed to test the statistical significant of both methods, and we arrived at the following conclusions:
\begin{enumerate}
  \item{DF method can correctly reproduce both the anisotropic amplitude and the anisotropic direction when the data are really anisotropic. The probability that DF method detects pseudo anisotropic signals when the data are actually isotropic is negligible. The inhomogeneous distribution of data points in the sky almost does not bias the final results. Therefore, the anisotropic signals found in Union2.1 using DF method is statistically significant.}
  \item{HC method couldn't correctly reproduce the anisotropic direction, although it can approximately reproduce the anisotropic amplitude. There is a high probability that HC method detects pseudo anisotropic signals when the data are actually isotropic. The homogeneous distribution of data points in the sky couldn't significantly improve this situation. Therefore, the preferred direction found using HC method is questionable.}
\end{enumerate}

Many reasons may lead to the low statistical significance of HC method. First, HC method depends on the separation of the sky. Due to the limited computational time, the sky is divided into limited number of pixels. HC method only compares the directions centering on each pixel, and skips all the other directions. However, dividing the sky into more pixels couldn't significantly improve the results. Second, HC method divides the sky into two opposite hemispheres, and compare the cosmological parameters in each hemisphere. The parameters derived in this way are the results of the average contribution from one hemisphere, but do not exactly represent the parameters in this specific direction. Third, and maybe more importantly, the redshifts of supernovae in Union2.1 are low. In such a low-redshift range, the distance modulus is insensitive to $\Omega_M$. Hence, a small noise of distance modulus may lead to a large fluctuation of $\Omega_M$. This problem could be solved by adding more high-redshift data such as gamma-ray bursts.

To alleviate the discrepancy between DF and HC methods, we proposed that the matter density can be modeled by the dipole form. By fitting our model to Union2.1, we obtained the direction of maximum $\Omega_M$ pointing towards $(l,b)=(351.8^{\circ}\pm 42.0^{\circ}, -9.9^{\circ}\pm 20.3^{\circ})$. This is well consistent with the dipole direction obtained using DF method, but it is about $75^{\circ}$ apart from the one obtained using HC method. This direction is also about $129^{\circ}$ away from the direction previously obtained using Union2 \citep{Antoniou:2010}. MC simulations shows that our result is statistically acceptable. The statistical significance can be further improved if the data points are homogeneously distributed in the sky.

The quality of supernovae data is not high enough. Especially, the sky coverage is not complete and the redshift coverage is very narrow. The anisotropic signals hiding in the supernovae depend on both the methods and data sets. Although the anisotropic signals found using DF method couldn't be completely contributed to statistical noise, the systematic uncertainty still cannot be excluded. It was shown that the anisotropic signals may completely disappear if the correlations between supernovae are considered \citep{Jimenez:2014}. Therefore, according to the present supernovae data, it is premature to claim that the Universe has any preferred direction.

\section*{Acknowledgements}
We are grateful to J. Li, H. Ma and L. Tang for useful discussions. X. Li has been supported by the National Natural Science Fund of China (NSFC) (Grant No. 11305181 and 11547305) and the Open Project Program of State Key Laboratory of Theoretical Physics, Institute of Theoretical Physics, Chinese Academy of Sciences, China (No. Y5KF181CJ1). Z. Chang has been funded by the NSFC under Grant No. 11375203.

\label{lastpage}

\end{document}